\documentclass[aps,prd,twocolumn,showkeys,superscriptaddress,groupedaddress,nofootinbib,floatfix]{revtex4}

\usepackage{graphicx}
\usepackage{dcolumn}
\usepackage{bm}

\usepackage{graphicx}	
\usepackage{amsmath}	
\usepackage{amssymb}	

\usepackage{epsfig,amsmath,natbib}
\usepackage{color,subfigure}
\usepackage{xcolor}

\usepackage{mathrsfs,amssymb,amstext}
\usepackage{ulem}
\usepackage{url}
\usepackage{bm}
\usepackage{adjustbox}
\usepackage{threeparttable}

\newcommand{\Mpc}{\mathrm{~km~s^{-1}~Mpc^{-1}}}

\begin{document}


\title{
Assessing the effect of mass-model assumptions on measuring the Hubble constant from the cluster-lensed supernova Refsdal}



\author{Yuting Liu\footnote{yutingl@imu.edu.cn}}
\affiliation{School of Physical Science and Technology, Inner Mongolia University, Hohhot 010021, China}

\author{Masamune Oguri\footnote{masamune.oguri@chiba-u.jp}} 
\affiliation{Center for Frontier Science, Chiba University,  Chiba 263-8522, Japan}
\affiliation{Department of Physics, Graduate School of Science, Chiba University, Chiba 263-8522, Japan}

\date{\today}

\begin{abstract}
The Hubble constant, $H_0$, which is a crucial parameter in astrophysics and cosmology, is under significant tension.
We explore an independent technique to measure $H_0$ based on the time-delay cosmography with strong gravitational lensing of a supernova lensed by a galaxy cluster, focusing on SN Refsdal in MACS J1149.5+2223, the first gravitationally lensed supernova with resolved multiple images. We carefully examine the dependence of constraints on the Hubble constant on the choice of lens mass models, employing 23 lens mass models with different assumptions on dark matter halos and external perturbations.
Remarkably, we observe that the dependence on the choice of lens mass models is not significantly large, suggesting the robustness of the constraint on the Hubble constant from SN Refsdal. We combine measurements for the 23 lens mass models to obtain $H_0=70.0^{+4.7}_{-4.9}\Mpc$ assuming equal weighting.  We find that best-fitting Hubble constant values correlate with radial density profiles of the lensing cluster, implying a room for improving the constraint on the Hubble constant with future observations of more multiple images. We also find a clear correlation between best-fitting Hubble constant values and magnification factors of supernova multiple images. This correlation highlights the importance of gravitationally lensed Type Ia supernovae for accurate and robust Hubble constant measurements. \end{abstract}

                 
\maketitle


\section{Introduction}\label{sec:intro}

The Hubble constant, denoted as $H_0$, quantifies the current expansion rate of the Universe and stands as a key parameter in modern cosmology. The Hubble constant is important, particularly because it is a cornerstone in determining fundamental properties such as the age, size, and fate of the Universe \cite{1929PNAS...15..168H,1931MNRAS..91..483L,2001ApJ...553...47F,2011Obs...131..394N,2013PhR...530...87W}.
Various methodologies have been employed to measure $H_0$. Early estimates relied on distance measurements to relatively nearby galaxies, while contemporary approaches incorporate diverse cosmological probes. These approaches include Type Ia supernova (SNe Ia), cosmic microwave background radiation (CMBR), and baryon acoustic oscillations (BAO), among others. However, despite significant observational progress, reconciling different measurements and reducing systematic uncertainties remain ongoing challenges. The quest to precisely determine $H_0$ has led to intriguing controversies
\cite{2019NatAs...3..891V,2021ApJ...919...16F,2022JHEAp..34...49A,2023Univ....9...94H,2023arXiv230109695L,2023arXiv231107552C}, which emerge when comparing local $H_0$ measurements with those inferred from observations of the early Universe. For instance, the distance ladder analysis from SH0ES (SNe, H0, for the Equation of State of dark energy) collaboration measures the Hubble constant of $H_0=73.04\pm 1.04 \Mpc$ in the late Universe \cite{2022ApJ...934L...7R}, while CMBR from the \textit{Planck} satellite releases a much smaller and tighter measurement with the value of $H_{0}=67.4\pm0.5\Mpc$ in the early Universe \cite{2020A&A...641A...6P}.  

The discrepancies of measured $H_{0}$ values between different measurements are becoming more pronounced.
Gravitational waves (GW) from the coalescence of binary systems are used as a new distance indicator that is independent of the local distance ladder, offering a unique technique to determine $H_0$. 
For instance, the neutron star merger event GW170817 coupled with its electromagnetic counterparts AT2017gfo and GRB170817A led to the Hubble constant of $H_0=66.2^{+4.4}_{-4.2}\Mpc$ \cite{2020Sci...370.1450D}. 
More recently, Ref.~\cite{2024MNRAS.tmp...74A} derived a Hubble constant value of  $H_0=68.84^{+15.51}_{-7.74}\Mpc$
from the first three LIGO/Virgo observing runs and DELVE, specifically events GW$190924{\_}021846$ and GW$200202{\_}154313$.
Analyzing the dispersion measures and redshifts of fast radio bursts (FRBs) is another approach to constrain the Hubble constant
\cite{2022MNRAS.509.4775J,2022MNRAS.511..662H,2022MNRAS.515L...1W}.
Ref.~\cite{2022MNRAS.516.4862J} employ such approach to derive
the value of $H_0=73^{+12}_{-8} \Mpc$, using a sample of 16
Australian Square Kilometre Array Pathfinder (ASKAP) FRBs detected
by the Commensal Real-time ASKAP Fast Transients (CRAFT) Survey and localised to their host galaxies, and 60 unlocalised FRBs from Parkes and ASKAP.
The tension observed among various measurements raises profound questions regarding the robustness of the standard cosmological model, or the potential existence of unknown physics influencing cosmic expansion. Therefore, exploring alternative possibilities is urgent to elucidate the origin of this tension and attain a concordant value for the Hubble constant.

Gravitational lensing time delay cosmography, initially proposed by Refsdal \cite{1964MNRAS.128..307R}, has evolved into a crucial independent method for constraining the Hubble constant, thanks to advancements in observations and theoretical understanding of the dependence on lens models.
Traditionally, constraints on $H_0$ from gravitational lens time delays make use of mainly galaxy-lensed quasar systems \cite{2019Sci...365.1134J,2019MNRAS.490.1743C,2017MNRAS.465.4895W,2019MNRAS.484.4726B,2020MNRAS.494.6072S,2017ApJ...838L..15L}.
Recently, Ref.~\cite{2023A&A...673A...9S} provide a new $H_0$ measurement ($H_0=73.3\pm5.8\Mpc$) from seven galaxy-lensed quasar systems using single-aperture kinematics and maximally flexible models. 

In addition to galaxy-lensed quasar systems, cluster-lensed quasar systems
have also been used to constrain the Hubble constant \cite{2021MNRAS.501..784D,2023arXiv230111240N,2023arXiv230914776M}. Even though cluster strong lens mass modeling exhibits relatively large positional offsets between observed and model-predicted multiple images, it is still possible to recover mass models accurately with help of the large number of multiple images \cite{2017MNRAS.472.3177M}.
Ref.~\cite{2023PhRvD.108h3532L} used
the first cluster-scale lensed quasar system SDSS J1004+4112 to obtain $H_0=67.5^{+14.5}_{-8.9}\Mpc$ combining sixteen different mass models, from which it is concluded that the lens model dependence is large and needs to be overcome.

Very recently, the Hubble constant has been derived from the cluster-lensed supernova system SN Refsdal in the field of MACS J1149.5+2223 which is the first reported case of strongly lensed supernovae with resolved multiple images \cite{2015Sci...347.1123K,2016ApJ...819L...8K}.
Focusing on this unique and valuable system, Ref.~\cite{2018ApJ...853L..31V} has carried out Refsdal’s original proposal to use a multiply imaged supernova to measure $H_0$, with the value of $H_0=64^{+9}_{-11}\Mpc$, in which preliminary values of time delay measurements from the early data in light curve observations have been used.
Ref.~\cite{2023Sci...380.1322K} measure
$H_0=66.6^{+4.1}_{-3.3}\Mpc$ based on accurate and precise measurements of time delays from the full light curve data.
Since their analysis is blinded in the sense that mass models used for the analysis are those constructed before the reappearance of the fifth image of SN Refsdal, they do not thoroughly analyze degeneracies between the lens models and the Hubble constant.
Furthermore, independent measurements of the Hubble constant and the geometry of the Universe from SN Refsdal is presented by Ref.~\cite{2024arXiv240110980G}, which argues that the mass model uncertainty is subdominant \cite{2020ApJ...898...87G}.

In this paper, we explore to establish an independent measurement of $H_0$ while meticulously examining the influence of the assumed lens mass model on $H_0$ from this cluster-lensed supernova system SN Refsdal within MACS J1149.5+2223.  Specifically, we incorporate more than 100 multiple images and the most recent time-delay measurements between distinct lensed-supernova images \cite{2016ApJ...819..114K,2023ApJ...948...93K}. Our work analyzes 23 different lens mass models with different complexity to thoroughly explore the dependence of assumptions on mass models on the derived $H_0$ values. Our approach  is complementary to Refs.~\cite{2023Sci...380.1322K} and \cite{2020ApJ...898...87G}, and offers a new test of the robustness of the Hubble constant measurements in previous analysis.
Additionally, we discuss shapes of critical curves, radial convergence, tangential shear profiles, and magnifications of lensed supernova images, in order to explore how the Hubble constant measurement can be improved in future observations.

This paper is structured as follows.
In Sec.~\ref{sec:data}, we provide a brief description of the cluster-lensed supernova system SN Refsdal.
Section ~\ref{sec:model} introduces the methodology employed for mass modeling. Results of our analysis are presented in Sec.~\ref{sec:result}.
Discussion and conclusions are presented in Sec.~\ref{sec:discussion} and Sec.~\ref{sec:summary}, respectively.
In this paper, we assume a flat Universe with $\Omega_{\rm m}=0.3$ and $\Omega_{\Lambda}=0.7$.

\section{Observations of SN Refsdal in MACS J1149.5+2223} \label{sec:data}

SN Refsdal as the first example of a strongly lensed supernova 
with resolved multiple images was initially detected in November 2014 through the Hubble Space Telescope (HST) observations \cite{2015Sci...347.1123K}.
This system is composed of six multiple images (named S1-S4, SX, and SY in \cite{2015MNRAS.449L..86O}) of a single core-collapse supernova at redshift 1.488, where four images S1–S4 forming an Einstein-cross configuration around an early-type galaxy were detected by the Grism Lens Amplified Survey from Space (GLASS; GO-13459; PI Treu) \cite{2014ApJ...782L..36S,2015ApJ...812..114T}, SX detected in HST imaging on 2015 December 11 (GO-14199; PI Kelly) \cite{2016ApJ...819L...8K} was the first supernova whose appearance on the sky was predicted in advance 
\cite{2015MNRAS.449L..86O,2015ApJ...800L..26S,2016MNRAS.456..356D}, 
and a leading image SY appeared in the late 1990s from the lens models predication without observation evidence.
SN Refsdal is lensed by the cluster of galaxies MACS J1149.5+2223 at redshift 0.541 \cite{2007ApJ...661L..33E}.  
Multiple images and their spectroscopic redshifts for this cluster have been confirmed and presented in the literature \cite{2009ApJ...703L.132Z,2009ApJ...707L.163S,2012Natur.489..406Z,
2014MNRAS.443..957R,2014MNRAS.444..268R,2016MNRAS.457.2029J,2016ApJ...817...60T,2016ApJ...822...78G,2016ApJS..226....6B}.

\begin{figure}
\includegraphics[width=0.96\linewidth]{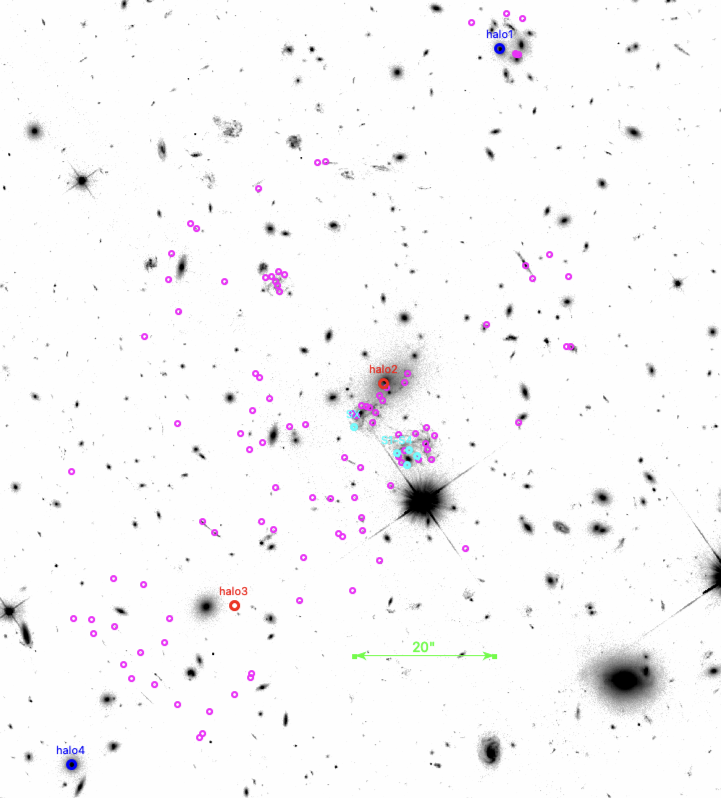}
\caption{The HST F606W image of MACS J1149.5+2223 with positions of multiple images of SN Refsdal ({\it cyan}) as well as multiple images of other background galaxies ({\it magenta}).  North is up and West is right. Positions of halo components are also indicated by large circles. For red circles, the positions are left as free parameters, while for blue circles the positions are fixed in  mass modeling. When including only 3 halos in mass modeling, 'halo 1' is not included. }

\label{ds9}
\end{figure}

In this study, we utilize the positions of the five supernova images in  Ref.~\cite{2023ApJ...948...93K}, which have been refined based on the Hubble Space Telescope (HST) Wide Field Camera 3 (WFC3) infrared (IR) detector F125W images acquired in 2015 and 2016. Additionally, we incorporate positional constraints derived from multiple images of seven knots within a lensed face-on spiral galaxy at redshift 1.488. For multiple images of other background galaxies, we include data from 28 multiple image systems compiled in Ref.~\cite{2016ApJ...819..114K}. The total number of multiple images in our study is 109, spanning across 36 systems. Fig.~\ref{ds9} shows the position of all the multiple images used for our analysis.
Our analysis includes newly measured time delays between lensed supernova images S2–SX and S1, as well as recently determined magnification ratios from the full light curve analysis \cite{2023ApJ...948...93K}. The magnification ratio S4/S1 is excluded due to compelling evidence ($>5\sigma$) indicating that image S4 has undergone chromatic microlensing \cite{2023Sci...380.1322K}. 

Table~\ref{table:1} provides a summary of observational constraints for multiple images of SN Refsdal. While Ref.~\cite{2023ApJ...948...93K} offers a detailed depiction of the likelihood distribution for both the time delay and magnification, along with the covariance matrix encompassing these measurements,
in our analysis we adopt a simplified approach by assuming symmetric Gaussian distributions for both time delay and magnification measurements without any covariance between measurements, mainly because of the limitation of the software \textsc{glafic} \cite{2010PASJ...62.1017O,2021PASP..133g4504O} that we use for our analysis. For instance, Ref.~\cite{2023ApJ...948...93K} report a relative time delay of $376.0^{+5.6}_{-5.5}$ days between images SX and S1, alongside a magnification ratio  SX/S1=$0.30^{+0.05}_{-0.03}$ (see Fig.14 and Fig.15 in \cite{2023ApJ...948...93K}). We will discuss a possible impact of this simplified assumption in Sec.~\ref{sec:result}.

Additionally, we incorporate information about the positions, ellipticities, position angles, and luminosity ratios (relative to the brightest cluster galaxy in the MACS J1149.5+2223 field) of 170 cluster galaxy members, as detailed in Ref.~\cite{2016ApJ...819..114K}. 
The large number of observational constraints suggests the potential for tightly constraining the Hubble constant through this lens system.

\begin{table}
  \caption{Observational constraints from multiple images of SN Refsdal.
  Source J2000 positions with respect to the (RA, DEC)=($177.3987491$, $22.3985308$), the magnification ratio with respect to the supernova image S1, and the relative time delay are listed in columns 2--5, respectively.
  \label{table:1}}
     \begin{threeparttable}
  \setlength{\tabcolsep}{2.6mm}{
    \begin{tabular}{ccccc}
     \hline\hline
      Name & $\Delta x$ [$''$] &  $\Delta y$[$''$]  & $\Delta \mu $ &
      $\Delta t$ [days]\\
     \hline
S1 &  1.754  & $-10.459$ &  $\equiv  1$   & $\equiv  0$ \\
S2 &  3.458  & $-9.908$ &  $1.06\pm0.25$ & $9.7\pm4.4$  \\
S3 &  4.596  & $-10.808$ &  $0.91\pm0.39$ & $7.9\pm5.6$  \\
S4 &  3.168  & $-12.068$ &  $\equiv  0 $ \tnote{*} & $19.4\pm8.0$\\
SX &  $-4.432$  & $-6.620$ &  $0.30\pm0.05$ & $376.0\pm5.6$   \\
  \hline\hline
\end{tabular}}
\begin{tablenotes}
\footnotesize
\item[*] {The magnification ratio S4/S1 is not
included because image S4 has experienced chromatic microlensing.}
\end{tablenotes}
\end{threeparttable}
\end{table}

\section{Mass Modeling} \label{sec:model}

We construct lens mass models utilizing the widely employed software \textsc{glafic} \cite{2010PASJ...62.1017O,2021PASP..133g4504O}, which adopts a parametric modeling approach.

\subsection{Dark matter halo}
We employ the Navarro-Frenk-White (NFW) profile to characterize the mass distribution of a dark matter halo \cite{1997ApJ...490..493N}. The projected density profile is defined by six parameters, including the mass $M$, the centroid position $x$ and $y$, the ellipticity $e$, the position angle $\theta_{\rm e}$, and the concentration parameter $c$. In our mass model reconstruction, we include up to four NFW components to model the complex cluster mass distribution. For NFW components located at the edge or outside the strong lensing regions, positions are fixed to observed positions of bright cluster member galaxies (see Fig.~\ref{ds9}). In cases where NFW components are located outside the main strong lensing region and lack strong observational constraints, the concentration parameter is fixed to 10, because the concentration parameter value is not very well determined from the data. Across all halo components, the halo mass and ellipticity are treated as free parameters. We optimize these model parameters, with the ellipticity restricted to be smaller than 0.8 and the concentration parameter restricted to the range between 1 and 40.
We utilize the fast approximation of the lensing calculation of the NFW profile, dubbed as {\tt anfw} in \textsc{glafic}, as proposed in Ref.~\cite{2021PASP..133g4504O}.

We also utilize the Pseudo-Jaffe Ellipsoid (PJE, dubbed as {\tt jaffe} in \textsc{glafic}) profile \cite{1983MNRAS.202..995J,2001astro.ph..2341K} to characterize the dark matter halo. In this model, the radial density profile is defined by seven parameters, including the velocity dispersion $\sigma$, the centroid position, the ellipicity, the position angle, the core radius $r_{\rm core}$, and the truncation radius $r_{\rm trun}$. As in the case of the NFW profile, their centroids positions are fixed when located at the edge or outside the strong lensing regions, while allowing all the other model parameters to be free during optimization. The optimization is conducted within the specified range of $r_{\rm trun}$ from $20''$ to $500''$ and $e$ smaller than $0.8$. 

Moreover, in some cases we simultaneously model the dark matter halo components with both the NFW and PJE profiles, employing similar construction and optimization strategies as outlined above.

\subsection{Member galaxies}

The member galaxies are modeled by scaled Pseudo-Jaffe ellipsoids (dubbed as {\tt gals} in \textsc{glafic}). This profile is parameterized by the scaled velocity dispersion $\sigma_*$, scaled truncation
radius $r_{\rm trun,*}$, and the scaling slope $\eta$. Throughout our optimization process, all these parameters are allowed to vary, with the range for $\eta$ from $0.2$ to $1.5$.
In addition, we separately model a cluster member galaxy that is responsible for producing the four Einstein cross supernova images S1-–S4, using the PJE profile, rather than the scaled Pseudo-Jaffe ellipsoids mentioned above. All model parameters for this member galaxy are treated as free parameters during optimization.

\subsection{Multipole perturbations}

In order to enhance the flexibility of our modeling, we introduce both an external shear to the lens potential and higher-order perturbations accounting for the potential asymmetry in the cluster mass distribution. These perturbations can significantly improve the mass modeling precision. To achieve this, we consider multipole perturbations $\phi=-(\epsilon/m)r^2\cos m(\theta-\theta_{\epsilon}-\pi/2)$ with orders up to 6, i.e., $m=2$, $3$, $4$, $5$, and
$6$ 
\cite{2003MNRAS.345.1351E,2004PASJ...56..253K,2005MNRAS.364.1459C,2006ApJ...642...22Y,2013MNRAS.429..482O}. In our optimization process, we optimize the model parameters $\epsilon$ and $\theta_{\epsilon}$  without imposing any prior ranges.
The external shear corresponds to the multipole perturbation with $m = 2$ \citep{1997ApJ...482..604K}, dubbed as {\tt pert} in \textsc{glafic}, while terms with $m\geq 3$ is dubbed as {\tt mpole} in \textsc{glafic}.

\subsection{Model optimization}

\begin{figure}
\includegraphics[width=0.96\linewidth]{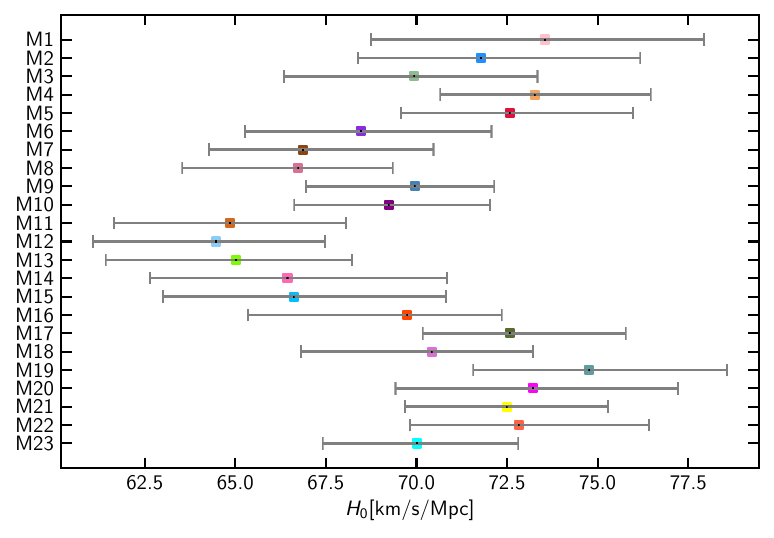}
\includegraphics[width=0.98\linewidth]{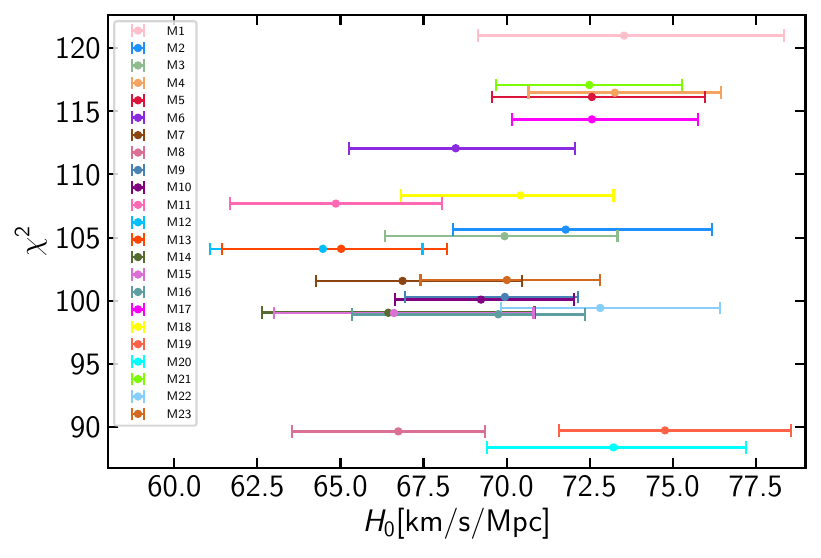}
\includegraphics[width=0.98\linewidth]{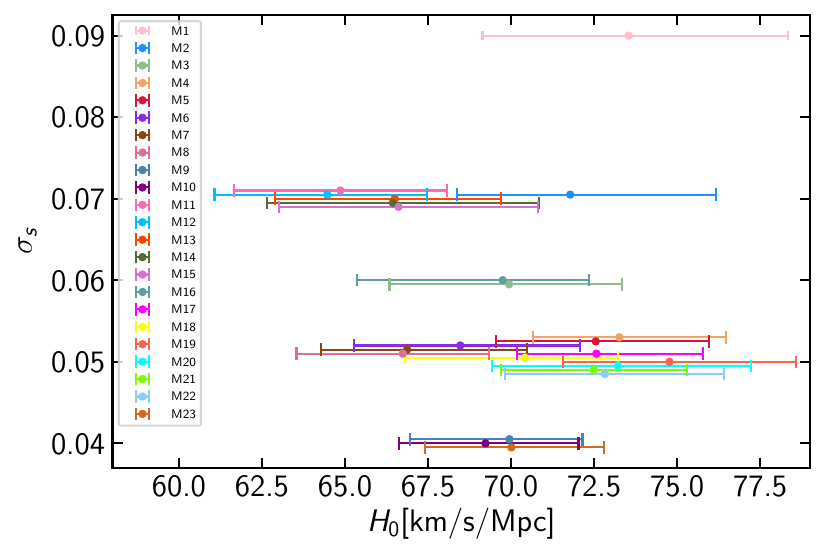}
\caption{{\it Top:} Summary of constraints on the Hubble constant across 23 different lens mass models.  {\it Middle:} Similar to the top panel, but presenting the correlation between $\chi^2$ values and measurements of the Hubble constant. {\it Bottom:} Similar to the top panel, but presenting the correlation between assumed position errors of supernova images (with small vertical offsets for clarity) with Hubble constant values. 
}
\label{Fig1}
\end{figure}

In this paper, we consider 23 different lens mass models to fit the cluster-lensed supernova system, aiming to explore the dependence of $H_0$ constraints on mass model assumptions as well as the dependency between $H_0$ and other physical properties associated with mass models. Specifically, we consider three variations in the utilization of different models to describe the dark matter halo, the number of dark matter halo components, and differences in the orders of perturbations.
More specifically, we utilize either the NFW profile ({\tt anfw} in \textsc{glafic}) or the PJE profile ({\tt jaffe} in \textsc{glafic}) to characterize the mass distribution of dark matter halos. Furthermore, we consider both cases of three dark matter halo components and four components. In terms of perturbations, we consider external shear ({\tt pert} in \textsc{glafic}) and multipole perturbations ({\tt mpole} in \textsc{glafic}) of order up to 6 at most. We expect that this comprehensive approach allows us to thoroughly explore the impact of different assumptions on the lens mass model and assess the robustness of previous measurements of $H_0$ from SN Refsdal \citep{2023Sci...380.1322K,2024arXiv240110980G}.

All our mass models are summarized in Table~\ref{table:2}, and are labeled from M1 to M23. For instance, model M1 employs the NFW profile ({\tt anfw}) to characterize three dark matter halo components, utilizes the scaled PJE profile ({\tt gals}) for most cluster members, and includes external shear ({\tt pert}). This model is denoted as M1 ({\tt anfw3}+{\tt gals}+{\tt pert}).  
In contrast, model M10 ({\tt anfw4}+{\tt gals}+{\tt pert}+{\tt mpole(m=3,4,5)}) features four dark matter halo components instead of three dark matter halo components in M1. Additionally, this model incorporates not only external shear but also multipole perturbations of order up to 5.
Furthermore, model M21 ({\tt anfw}+{\tt anfw3}+{\tt gals}+{\tt pert}+{\tt mpole(m=3)}) signifies the consideration of four dark matter halos with external shear and third-pole perturbation. However, this model incorporates the NFW profile for one dark matter halo component while employing the PJE profile for the others. We note that the Oguri-a* model in Ref.~\cite{2023Sci...380.1322K} corresponds to M7 in Table~\ref{table:2}.

We follow the standard mass modeling procedure by assuming positional errors that are larger than measurement errors. This larger positional error accounts for the complexities of cluster mass distributions, such as asymmetry and substructures, which may not be fully captured in our parametric mass modeling approach. Our choice of positional errors for each mass model is guided by the aim to achieve a reduced-$\chi^2$ value for the best-fitting model that is approximately equal to one.
Following \cite{2015MNRAS.449L..86O} and \cite{2016ApJ...819L...8K}, we assign smaller positional errors for the core and knots of the lensed spiral galaxy, $\sigma_{\rm h}$, compared to multiple images of  other background galaxies, $\sigma_{\rm g}$. Furthermore, we assign even smaller errors for the supernova images, recognizing the importance of accurately reproducing supernova image positions for precise predictions of time delays between multiple images \cite{2019MNRAS.489.2097B}. Specific values of positional errors adopted for the 23 different lens mass models are summarized in Table~\ref{table:2}.

In our mass modeling, we include a total of 109 multiple images originating from 36 systems, resulting in 225 observational constraints including magnification ratios and time delays for SN Refsdal (see Table~\ref{table:1}). Throughout all calculations and model parameter optimizations, we employ a standard  $\chi^2$ minimization method implemented in the \textsc{glafic} software. To enhance computational efficiency, we estimate $\chi^2$ in the source plane \cite{2010PASJ...62.1017O}.

\section{Constraints on Hubble constant}\label{sec:result}
\begin{figure}
\includegraphics[width=0.97\linewidth]{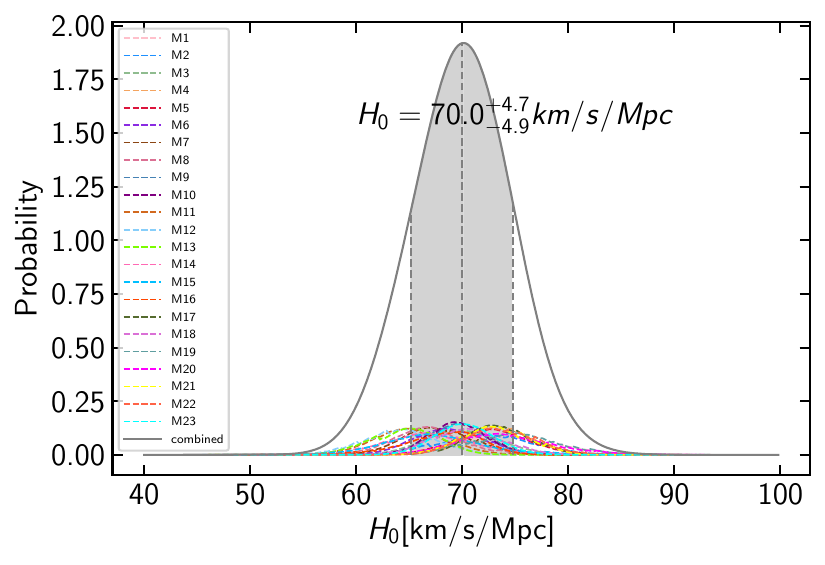}
\includegraphics[width=0.97\linewidth]{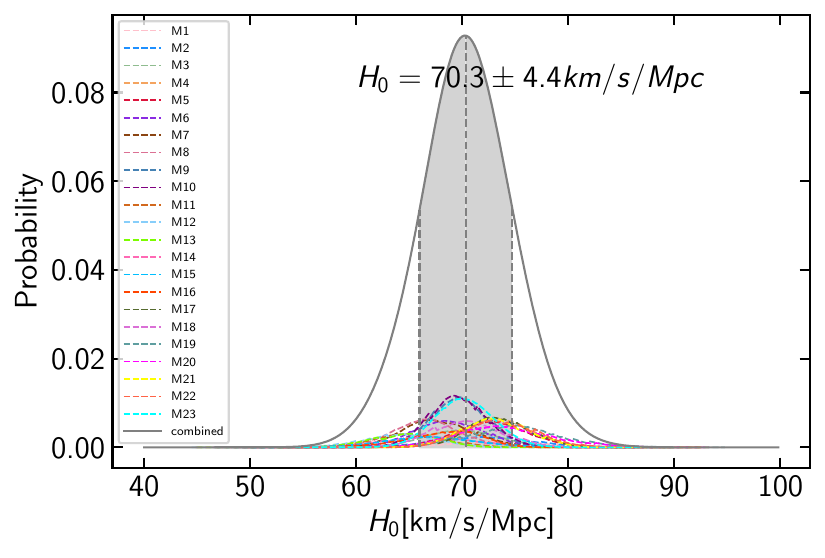}
\caption{{\it Top:} The probability distribution functions (PDFs) of the Hubble constant for all 23 different lens mass models ({\it dashed}), along with the PDF after combining all 23 mass models with equal weighting ({\it solid}). The vertical gray dotted lines and shaded region denote the median and $68.3\%$ confidence interval for the PDF of the combined result.
{\it Bottom:} Similar to the top panel, but combining all mass models with weighting based on the assumed positional errors of the supernovae images.
}
\label{Fig2}
\end{figure}

\begin{table*}
  \caption{Summary of assumed positional errors of supernovae images $\sigma_{\rm s}$, host knots images $\sigma_{\rm h}$, and galaxy images $\sigma_{\rm g}$, as well as the Hubble constant with 68.3$\%$ confidence interval and reduced-$\chi^2$ ($\chi^2$/dof) values for 23 different lens mass models.} 
  \label{table:2}
  \begin{threeparttable}
  \setlength{\tabcolsep}{3.0mm}{
    \begin{tabular}{ccccccc}
     \hline\hline
Label &  Model \tnote{*}    &  $\sigma_{\rm s}$[$''$]   &  $\sigma_{\rm h}$[$''$]   & $\sigma_{\rm g}$[$''$] & $H_0$ [$\Mpc$]& $\chi^2$/dof  \\\\
    \hline
M1 &{\tt anfw3}+{\tt gals}+{\tt pert}  &0.09 & 0.36 & 0.72 & $73.5 ^{+4.8}_{-4.4}$ &120.99/108\\
M2 &{\tt anfw3}+{\tt gals}+{\tt pert}+{\tt mpole}(m=3)  &0.07 & 0.28 & 0.56  & $71.8^{+4.4}_{-3.4}$ &105.63/106)\\
M3 &{\tt anfw3}+{\tt gals}+{\tt pert}+{\tt mpole}(m=3,4)   &0.06 & 0.24 & 0.48 & $69.9^{+3.4}_{-3.6}$ &105.11/104\\
M4 &{\tt anfw3}+{\tt gals}+{\tt pert}+{\tt mpole}(m=3,4,5) &0.05 & 0.20 & 0.40 & $73.3^{+3.2}_{-2.6}$ &116.47/102\\ 
M5 &{\tt anfw3}+{\tt gals}+{\tt pert}+{\tt mpole}(m=3,4,5,6) &0.05 & 0.20 & 0.40 & $72.6^{+3.4}_{-3.0}$ &116.13/100\\ 
   \hline
M6 &{\tt anfw4}+{\tt gals}+{\tt pert}  &0.05 & 0.20 & 0.40  & $68.5^{+3.6}_{-3.2}$ &112.07/104\\
M7 &{\tt anfw4}+{\tt gals}+{\tt pert}+{\tt mpole}(m=3)  &0.05 & 0.20 & 0.40  & $66.9^{+3.6}_{-2.6}$ &101.57/102 \\
M8 &{\tt anfw4}+{\tt gals}+{\tt pert}+{\tt mpole}(m=3,4)   &0.05 & 0.20 & 0.40  & $66.7^{+2.6}_{-3.2}$ &89.66/100\\
M9 &{\tt anfw4}+{\tt gals}+{\tt pert}+{\tt mpole}(m=3,4,5) &0.04 & 0.16 & 0.32 & $70.0^{+2.2}_{-3.0}$ &100.30/98\\ 
M10 &{\tt anfw4}+{\tt gals}+{\tt pert}+{\tt mpole}(m=3,4,5,6) &0.04 & 0.16 & 0.32 & $69.2^{+2.8}_{-2.6}$ &100.09/96\\ 
   \hline
M11 &{\tt jaffe3}+{\tt gals}+{\tt pert}  &0.07 & 0.28 & 0.56 & $64.9^{+3.2}_{-3.2}$  &107.69/104\\
M12&{\tt jaffe3}+{\tt gals}+{\tt pert}+{\tt mpole}(m=3)  &0.07 & 0.28 & 0.56 & $64.5^{+3.0}_{-3.4}$ &104.11/102 \\
M13&{\tt jaffe3}+{\tt gals}+{\tt pert}+{\tt mpole}(m=3,4)  &0.07 & 0.28 & 0.56& $65.0^{+3.2}_{-3.6}$ &100.86/100 \\
M14&{\tt jaffe3}+{\tt gals}+{\tt pert}+{\tt mpole}(m=3,4,5)&0.07 & 0.28 & 0.56 & $66.4^{+4.4}_{-3.8}$ &99.05/98\\ 
M15&{\tt jaffe3}+{\tt gals}+{\tt pert}+{\tt mpole}(m=3,4,5,6) &0.07 & 0.28 & 0.56& $66.6^{+4.2}_{-3.6}$ &99.03/96\\ 
   \hline
M16&{\tt jaffe4}+{\tt gals}+{\tt pert}  &0.06 & 0.24 & 0.48 & $69.8^{+2.6}_{-4.4}$  &98.93/99\\
M17&{\tt jaffe4}+{\tt gals}+{\tt pert}+{\tt mpole}(m=3)  &0.05 & 0.20 & 0.40 & $72.6^{+3.2}_{-2.4}$ &114.36/97\\
M18&{\tt jaffe4}+{\tt gals}+{\tt pert}+{\tt mpole}(m=3,4)  &0.05 & 0.20 & 0.40 & $70.4^{+2.8}_{-3.6}$ &108.34/95\\
M19&{\tt jaffe4}+{\tt gals}+{\tt pert}+{\tt mpole}(m=3,4,5) &0.05 & 0.20 & 0.40 & $74.8^{+3.8}_{-3.2}$ &89.74/93\\ 
M20&{\tt jaffe4}+{\tt gals}+{\tt pert}+{\tt mpole}(m=3,4,5,6) &0.05 & 0.20 & 0.40& $73.2^{+4.0}_{-3.8}$ &88.39/91\\
   \hline
M21&{\tt anfw}+{\tt jaffe3}+ {\tt gals}+{\tt pert}+{\tt mpole}(m=3)  &0.05 & 0.20 & 0.40 & $72.5^{+2.8}_{-2.8}$  &117.08/99\\
M22&{\tt anfw2}+{\tt jaffe2}+ {\tt gals}+{\tt pert}+{\tt mpole}(m=3)&0.05 & 0.20 & 0.40 & $72.8^{+3.6}_{-3.0}$ &99.43/100\\ 
M23&{\tt anfw3}+{\tt jaffe1}+ {\tt gals}+{\tt pert}+{\tt mpole}(m=3) &0.04 & 0.16 & 0.32 & $70.0^{+2.8}_{-2.6}$ &101.64/101\\
  \hline
  \hline
\end{tabular}}
\begin{tablenotes}
\footnotesize
\item[*] {For model components, {\tt anfw} denotes the NFW profile for the halo, {\tt jaffe} denotes the PJE profile for the halo, {\tt gals} denotes the scaled PJE profile for
most member galaxies, {\tt pert} refers to an external shear, and {\tt pert} refers to multipole perturbations of order m. The numerical value following {\tt anfw} or {\tt jaffe} indicates the number of halos. Please refer to the main text for a detailed explanation.}
\end{tablenotes}
\end{threeparttable}
\end{table*}

\begin{figure*}
\centering
\includegraphics[width=8.4cm]{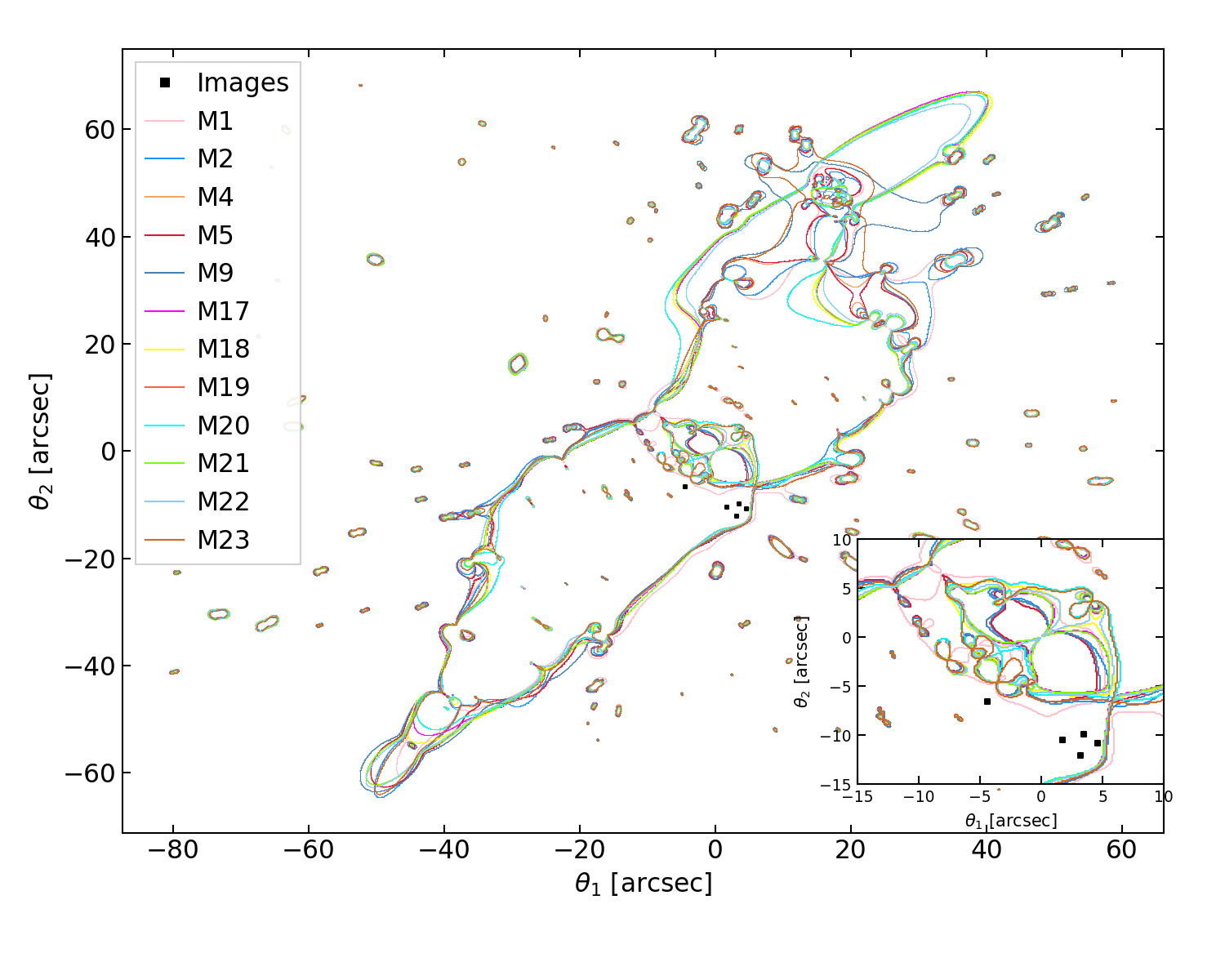}
\includegraphics[width=8.4cm]{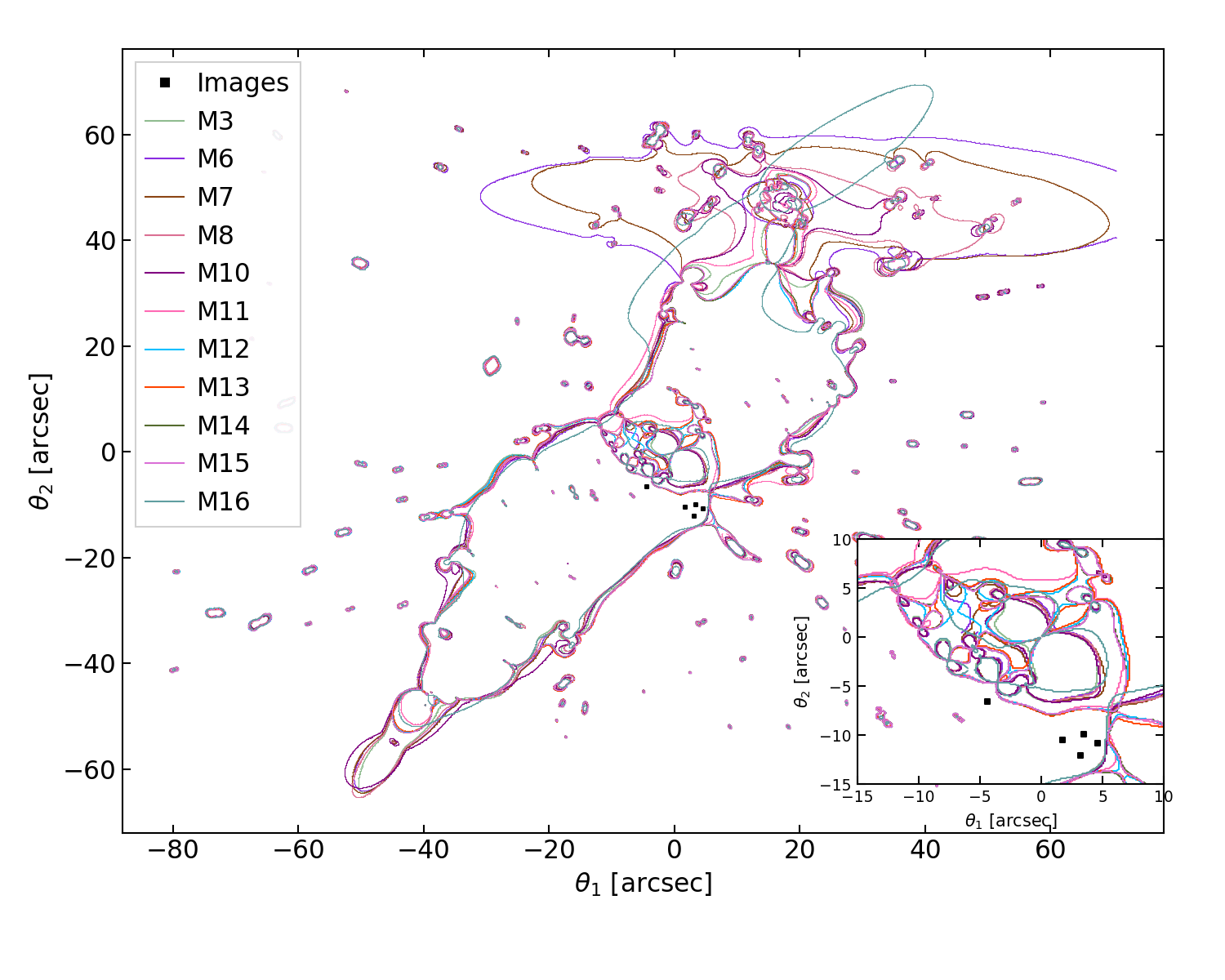}
\caption{
Critical curves for the 23 different lens mass models, 
computed assuming the source redshift of $z_{\rm s} = 6$. 
{\it Left:} Critical curves for 12 different lens mass models with best-fitting values of the Hubble constant larger than $70.0\Mpc$.  {\it Right:} Critical curves for 11 different lens mass models with best-fitting values of the Hubble constant smaller than $70.0\Mpc$. Black squares represent lensed supernova images. The lower right of each panel shows more details around the lensed supernova images.
}
\label{Fig3}
\end{figure*}

\begin{figure*}
\includegraphics[width=8.4cm]{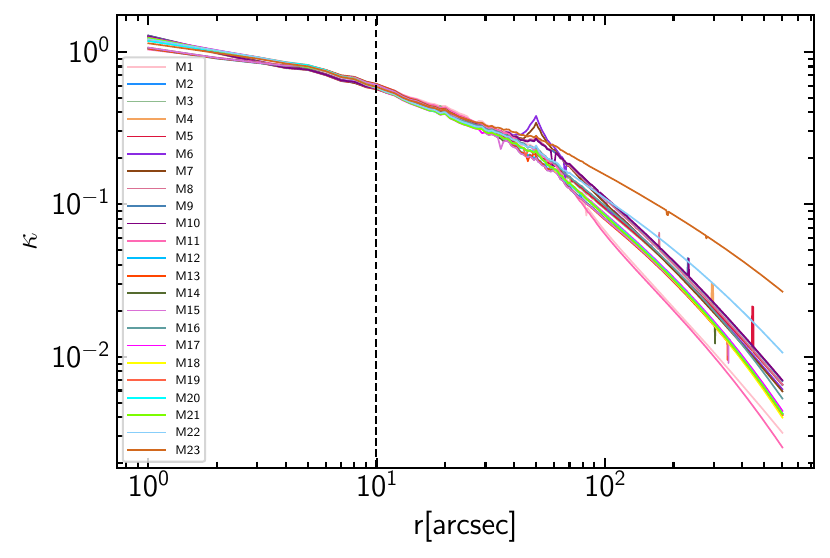}
\includegraphics[width=8.4cm]{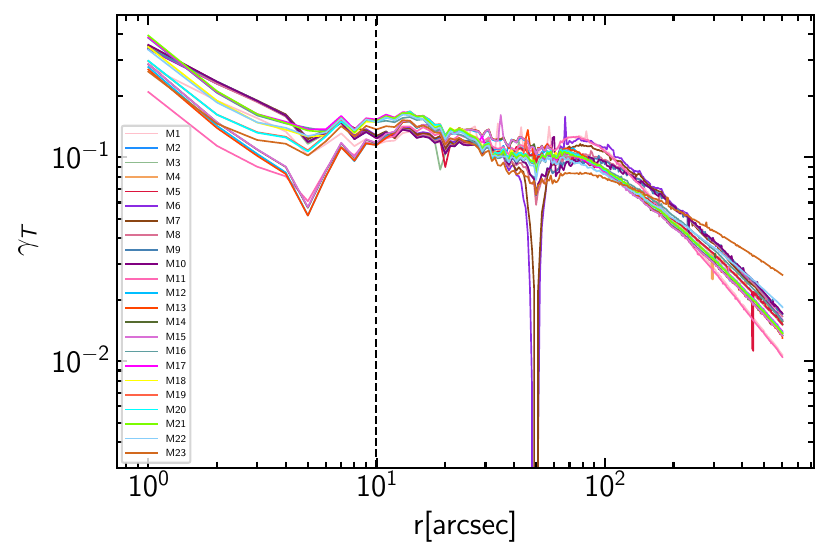}
\caption{Radial convergence ({\it left}) and tangential
shear ({\it right}) profiles for 23 different lens mass models, computed assuming the source redshift of $z_{\rm s}=1$. The Einstein radius for the source redshift of SN Refsdal is displayed by black vertical dashed lines.
}
\label{Fig4}
\end{figure*}

\begin{figure*}
\includegraphics[width=8.4cm]{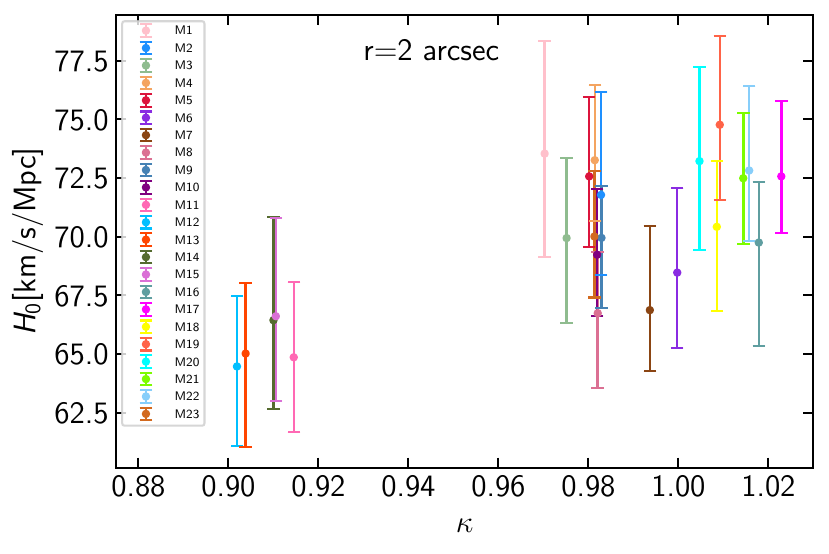}
\includegraphics[width=8.4cm]{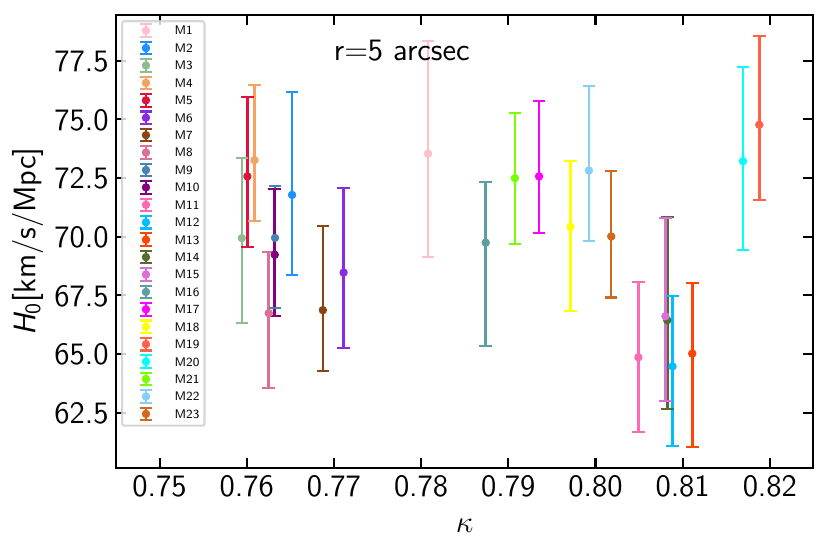}
\includegraphics[width=8.4cm]{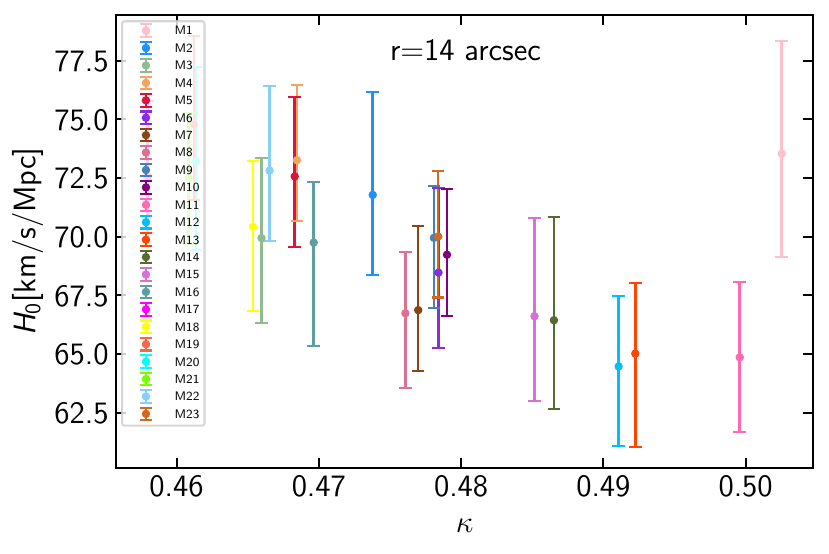}
\includegraphics[width=8.4cm]{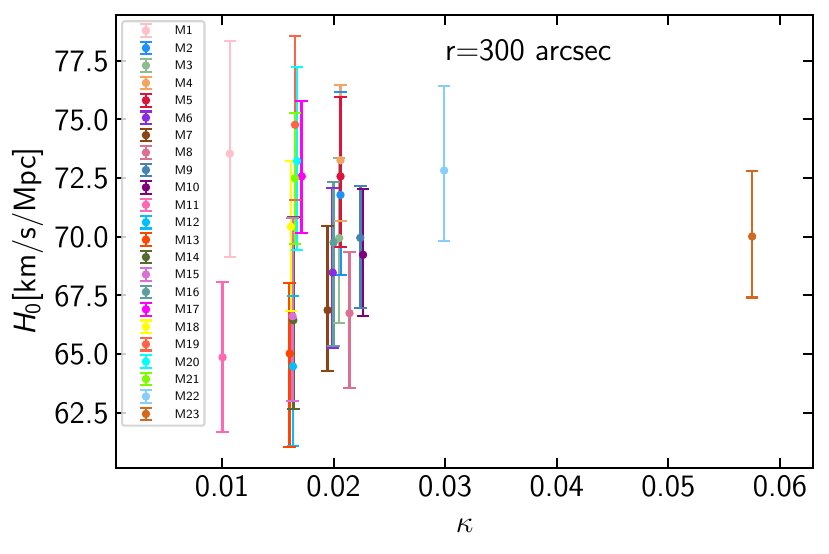}
\caption{The correlation between radial convergence values $\kappa$ at specific radius ($r=2{''}, 5{''}, 14{''},  300{''})$ and Hubble constant measurements for 23 different lens mass models.}
\label{Fig5}
\end{figure*}

\begin{figure*}
\includegraphics[width=8.4cm]{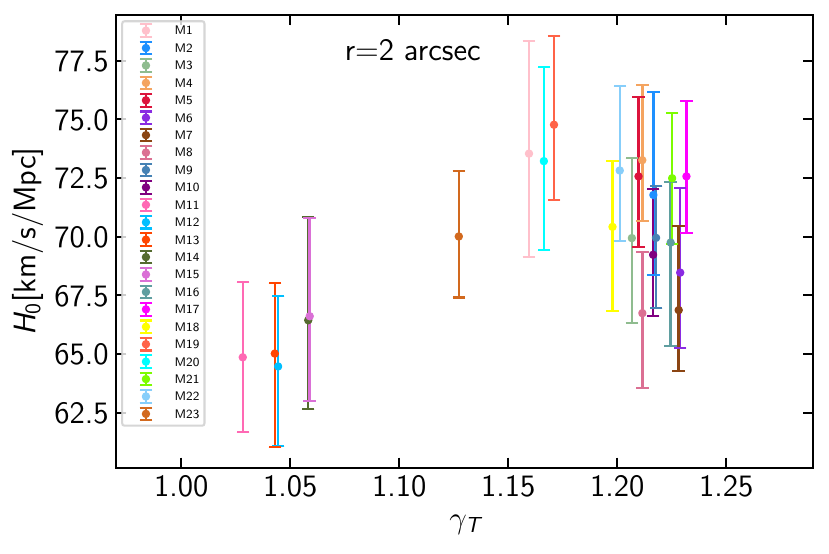}
\includegraphics[width=8.4cm]{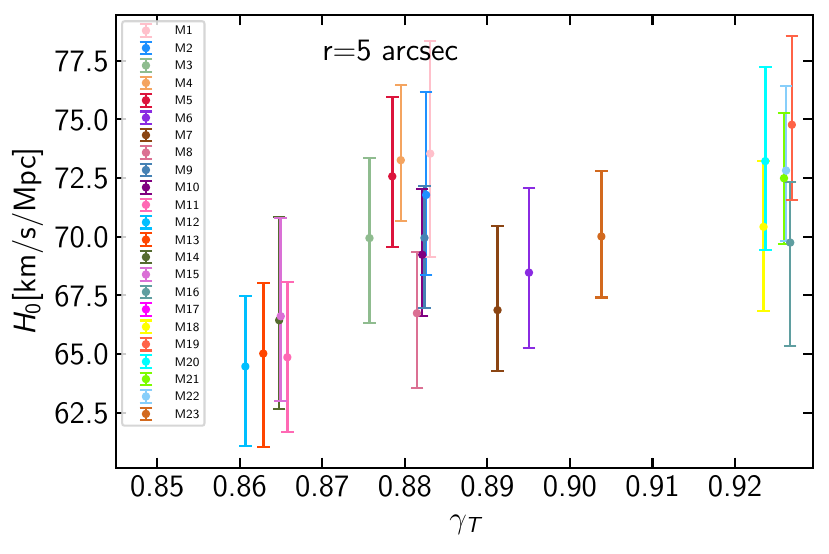}
\includegraphics[width=8.4cm]{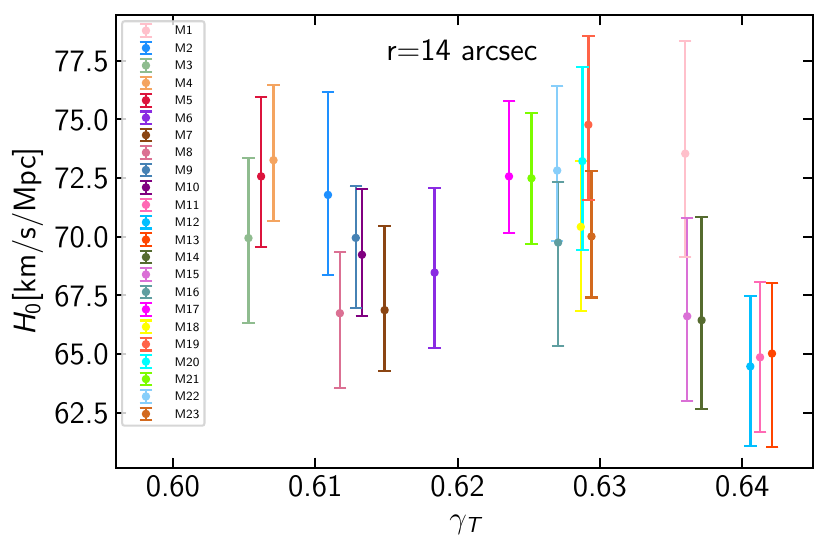}
\includegraphics[width=8.4cm]{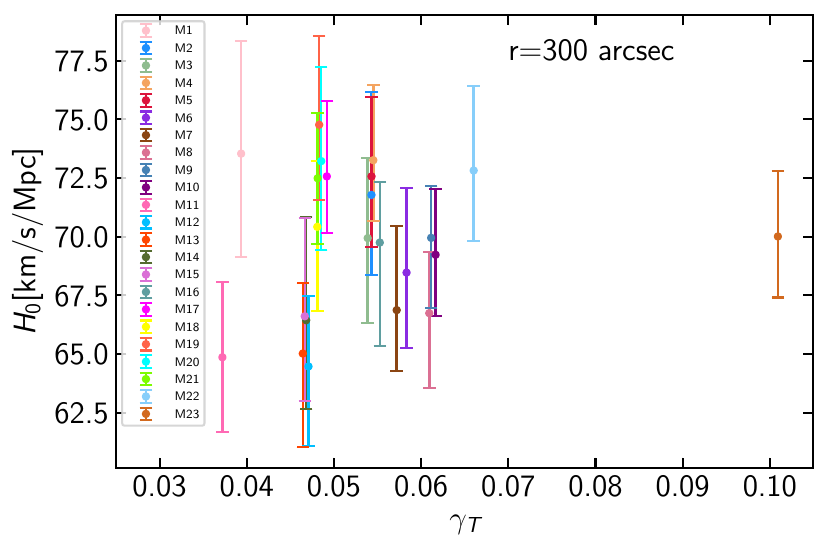}
\caption{Similar to Fig.~\ref{Fig5}, but for the tangential shear $\gamma_{\rm T}$.}
\label{Fig6}
\end{figure*}

\begin{figure*}
\includegraphics[width=8.4cm]{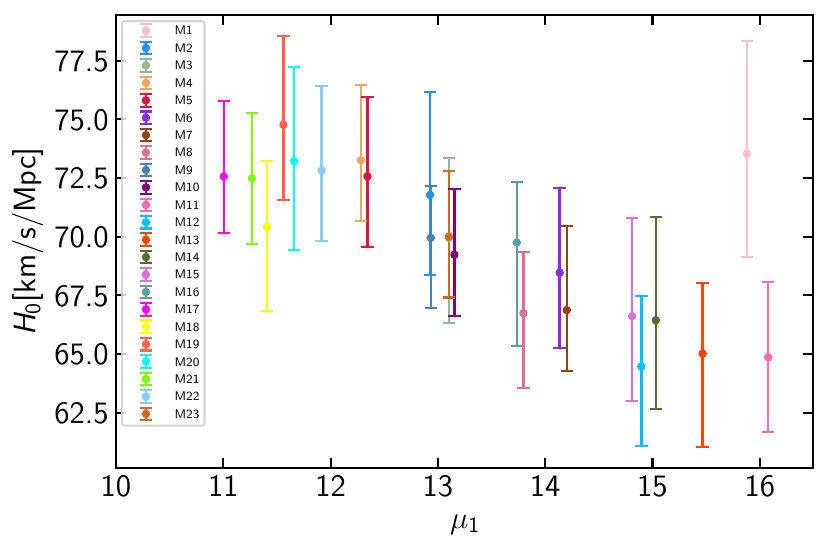}
\includegraphics[width=8.4cm]{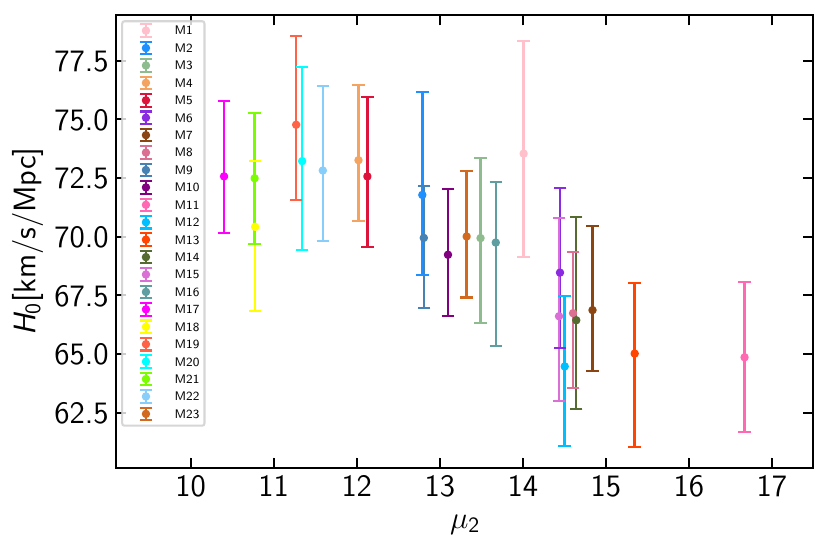}
\includegraphics[width=8.4cm]{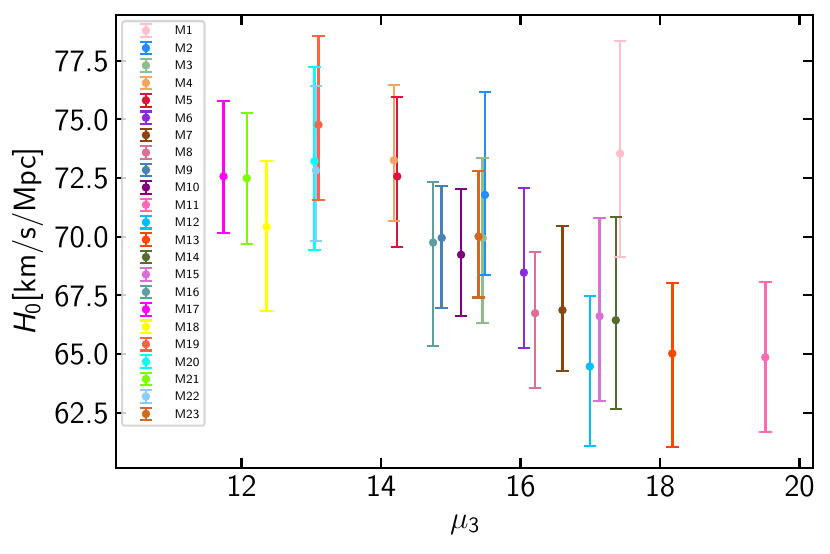}
\includegraphics[width=8.4cm]{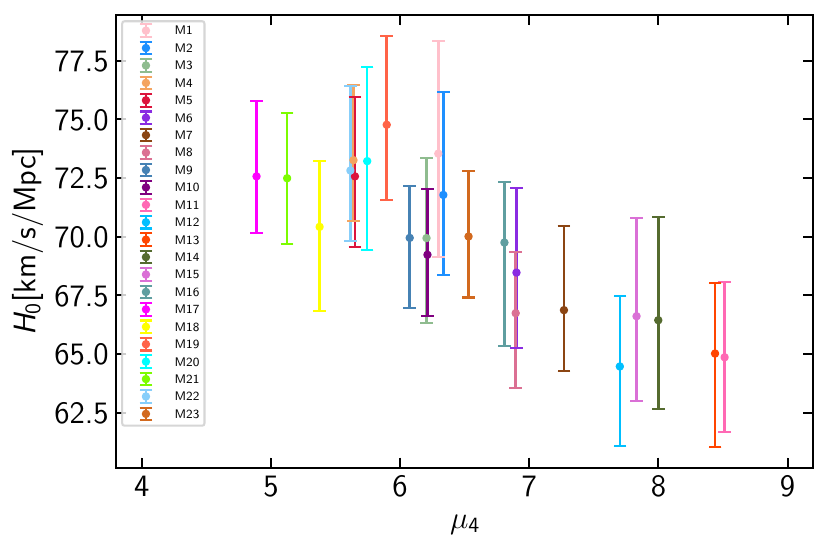}
\includegraphics[width=8.4cm]{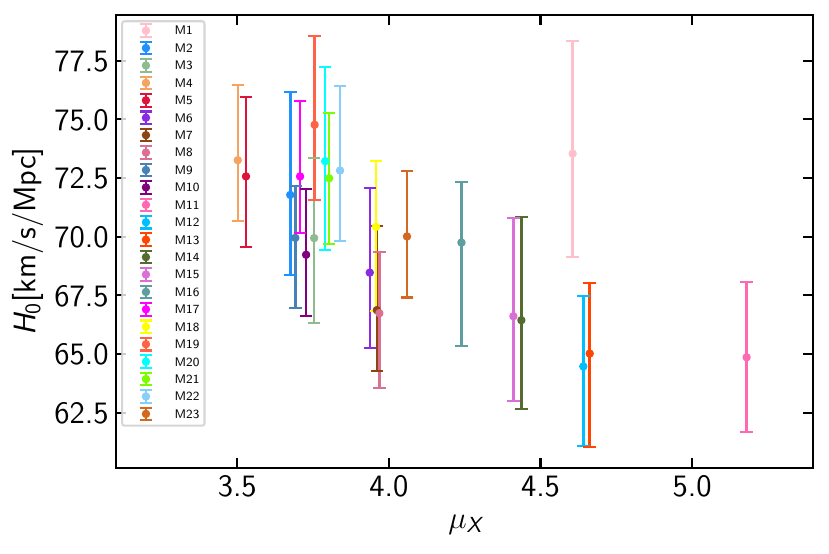}
\caption{The correlation between magnification factors of five multiple supernova images and Hubble constant measurements for 23 different lens mass models.}
\label{Fig7}
\end{figure*}

In the realm of time-delay cosmography, the Hubble constant $H_0$ can be directly inferred from time-delay measurements  $\Delta t$ and well-reconstructed lens potential $\phi$.
In our analytical framework, we incorporate $H_0$ as a model parameter, varying it in conjunction with all the other model parameters to simultaneously optimize the alignment of image positions, magnification ratios, and time delays, as detailed in Table~\ref{table:1}.
The uncertainty associated with $H_0$ is calculated by systematically varying its value around the best-fitting estimate, utilizing a step size of $0.2\Mpc$.
Specifically, 68.3\% confidence interval is derived from 
the range where the difference of the best-fitting $\chi^2$, denoted as $\Delta\chi^2$, is less than $1$, under the assumption of Gaussian errors.

We present our findings in Table~\ref{table:2} and Fig.~\ref{Fig1}. We observe slight differences in the best-fitting Hubble constant values across different lens mass models. This result is in marked contrast to our previous study in Ref.~\cite{2023PhRvD.108h3532L} focusing on the cluster-lensed quasar system SDSS J1004+4112, in which large variations of Hubble constant measurements across different mass models are observed. For instance, when employing the NFW profile to model the dark matter halo and incorporating external shear to account for perturbations (M1), we derive $H_0=73.5^{+4.8}_{-4.4}\Mpc$. Upon introducing additional multipole perturbations, the estimated $H_0$ values become $H_0=71.8^{+4.4}_{-3.4}\Mpc$  for M2, and $H_0=72.6^{+3.4}_{-3.0}\Mpc$ for M5. Notably, we observe minor differences upon adding more multipole perturbations across all considered mass models, as depicted in the top panel of Fig.~\ref{Fig1}.

We find that changing the number of dark matter halo components, particularly from three to four, has a modest impact on the resulting $H_0$ value. 
For instance, within the lens modeling combination M7, we obtain a Hubble constant value of $H_0=66.9^{+3.6}_{-2.6}\Mpc$. This value is smaller than that derived from the M2 model. 
Furthermore, changing the mass distribution modeling of the dark matter halo from the NFW profile to the PJE profile also influences the results. For instance, $H_0=69.2^{+2.8}_{-2.6}\Mpc$ is derived from M10, which is smaller than the $H_0$  value obtained for M20. 
Additionally, when incorporating four dark matter halo components with external shear and third-order perturbation, but utilizing the NFW profile for two dark matter halo components and the PJE profile for the others, we derive a Hubble constant constraint of $H_0=72.8^{+3.6}_{-3.0}\Mpc$. 
In general, we do not observe significant lens model dependence comparable to Ref.~\cite{2023PhRvD.108h3532L}, which examined the cluster-lensed quasar system SDSS J1004+4112. The difference between SN Refsdal and SDSS J1004+4112 presumably originates from the difference in the number of multiple images. In the case of SN Refsdal, the cluster mass distribution is tightly constrained by more than 100 multiple images, leading to the robust Hubble constant value against difference choice of mass models.

The Hubble constant value of $H_0=66.9^{+3.6}_{-2.6}\Mpc$ for M7 should be compared with the Hubble constant value of $H_0=66.6^{+4.1}_{-3.3}\Mpc$ from the combination of the Oguri-a* model (higher weight) and Grillo-g model (lower weight) in Ref.~\cite{2023Sci...380.1322K}. The very similar best-fitting values between these result basically support the validity of our analysis. We find that the M7 result exhibits a slightly tighter constraint, which may partly arise from our simplified assumption of Gaussian distributions without incorporating any covariance between measurements of magnifications and time delays. Another reason may be that the latter constraint contains the contribution from the Grillo-g model, albeit with the smaller weight. As it will be shown below, the uncertainty of our combined constraint on the Hubble constant largely comes from the variation of the best-fitting Hubble constant values among different models, rather than statistical errors for individual mass models, we expect that the impact of our simplified assumption on the final result is relatively minor.

To integrate findings from various lens mass models, we  follow the approach in Ref.~\cite{2023PhRvD.108h3532L} to aggregate the posteriors $\exp(-\Delta\chi^2/2)$ of the Hubble constant from individual mass models with equal weighting. The resultant combined posterior probability distribution function (PDF) of $H_0$, alongside PDFs for individual lens mass models, is depicted in Fig.~\ref{Fig2}. The constraint on $H_0$ from this combined PDF is $H_0=70.0^{+4.7}_{-4.9}\Mpc$ which provides a much tighter constraint compared to $H_0=67.5^{+14.5}_{-8.9}\Mpc$ derived from Ref.~\cite{2023PhRvD.108h3532L}.

To explore the impact of weighting on the result, we also combine all the lens mass models using weighting based on the assumed positional errors of supernovae images $\sigma_{\rm s}$. Specifically, we adopt the weighting value of $w_{ i}=1/{\sigma^{2}_{\rm s}}_{,i}$, where ${\sigma_{\rm s}}_{,i}$ represents the positional errors of individual mass models as listed in Table~\ref{table:2}. With this weighting scheme, we obtain $H_0=70.3\pm4.4\Mpc$ (refer to Fig.~\ref{Fig2}), representing a slightly tighter, but quite consistent, constraint compared to the equal weighting case.

Our analysis indicates that there is no significant dependence of the Hubble constant on assumptions on lens mass models, and the uncertainty from lens mass model assumptions is not unsatisfactory large compared with current observational constraints on the Hubble constant with SN Refsdal, which is consistent with Ref.~\cite{2020ApJ...898...87G}.

\section{Discussion}\label{sec:discussion}

\subsection{Critical curves}
The shape and extent of critical curves are highly sensitive to the mass distribution of the lensing objects. Therefore, we begin, by checking critical curves, to examine potential dependence of Hubble constant values with cluster mass distributions.
The results are displayed in Fig.~\ref{Fig3}. 
We find that there are some notable differences of critical curves, especially in the northern part. These differences arise largely from the consideration of varying numbers of dark matter halo components. 
The large variation of critical curves in the northern part, however, is seen in both models with large and small values of the Hubble constant.
This means that the uncertainty of the northern structure of the critical curve on the Hubble constant is not very significant. 
To summarize, our analysis reveals no significant correlations between critical curve shapes and the Hubble constant constraints across all examined lens mass models.

\subsection{Radial convergence and tangential shear profiles}

Next, we calculate the azimuthal average of the convergence at each radius to derive radial convergence and tangential shear profiles around the main halo ('halo 2' in Fig.~\ref{ds9}) for all the 23 different lens mass models. 
The radial convergence, denoted by $\kappa$, is directly related to the surface mass density of the lensing object and provides insights into its overall mass distribution.
On the other hand, the tangential shear, denoted by $\gamma_{\rm T}$, measures the coherent distortion of background sources around the lensing object. Unlike the radial convergence, the tangential shear describes the anisotropic stretching of images along the tangential direction with respect to the lens center. The tangential shear is sensitive to the gradient of the lens potential and offers valuable information about the internal structure and orientation of the lensing mass distribution.

Fig.~\ref{Fig4} illustrates the convergence and the tangential shear profiles for all the 23 different lens mass models. While there is a general similarity among the profiles across different lens mass models, some differences still persist. Therefore, we concentrate on specific radii ($r=2{''}, 5{''}, 14{''}, 300{''} $) to check the correlation between the Hubble constant and the convergence and the tangential shear values. 
We present our results in Fig.~\ref{Fig5} and Fig.~\ref{Fig6}.
Intriguingly, we observe that best-fitting Hubble constant values tend to increase with $\kappa$ and $\gamma_{\rm T}$ in the inner regions, while they decrease in the outer regions.

Our result indicates that best-fitting Hubble constant values are larger for models with steeper radial density profiles. It has been known that time delays sensitively depend on the radial density profile of the lens such that steeper profiles result in larger time delays. (e.g., \cite{1995ApJ...443...18W,2002ApJ...578...25K}), and our result is in line with those previous findings. In the case of SN Refsdal, the radial density profile is tightly constrained by many multiple images with different source redshifts, yet such correlation between the radial density profile and the Hubble constant is seen. This result implies that finding more multiple images in future deep observations can improve the constraint on the Hubble constant from SN Refsdal.

\subsection{Magnification factors}
Ref.~\cite{2003MNRAS.338L..25O} proposed that the measurement of magnification factors of lensed supernova images, which is available when the source is Type Ia supernova, breaks the degeneracy between the radial density profile and the Hubble constant and significantly improves the constraint on the Hubble constant, because magnification factors are sensitive to the radial density profile. It is however not immediately clear whether this method is indeed useful for massive clusters for which the radial profile is already constrained by many multiple images and also their mass distributions are highly complex.  Therefore, it is of great interest to explicitly check whether there is a correlation between best-fitting Hubble constant values and the magnification factors for five multiple images of SN Refsdal. We caution that SN Refsdal is a core-collapse supernova and hence their magnification factors have not been measured. As a result, in our measurements of the Hubble constant with SN Refsdal, any information on the magnification factors are not used. Our purpose here is to check the potential usefulness of lensed Type Ia supernovae in future analysis.

Our result shown in Fig.~\ref{Fig7} indicates that there is a clear correlation between Hubble constant values and magnification factors, well demonstrating the usefulness of magnification factor measurements \cite{2003MNRAS.338L..25O}, even for massive clusters.
Our analysis highlights the importance of gravitationally lensed Type Ia supernova, such as the recently discovered Supernova H0pe in the galaxy cluster PLCK G165.7+67.0 \cite{2024ApJ...961..171F}, for precise and accurate measurements of the Hubble constant.

\section{Conclusion}\label{sec:summary}

In this paper, we investigate the power of constraining the Hubble constant through time-delay cosmography utilizing cluster-lensed supernova systems. As a specific example, we focus our attention on the strongly lensed supernova system SN Refsdal in MACS J1149.5+2223, which represents the first gravitationally lensed supernova with resolved multiple images. The background supernova is lensed into six distinct multiple images. Additionally, many multiple images of galaxies located behind the lensing cluster have been identified. In total, our analysis encompasses 109 multiple images from 36 systems.

To explore the dependence of constraints on the Hubble constant on the choice of lens mass models, we utilize 23 different lens mass models, each making different assumptions regarding the profiles of the dark matter halo, the number of dark matter halo components, and multipole perturbations. We find that the variation in the best-fitting values of the Hubble constant is relatively small across the 23 lens mass models. By summing the posteriors of the Hubble constant with equal weighting, we obtain a combined constraint of $H_0=70.0^{+4.7}_{-4.9}\Mpc$. We  also try a different weighting scheme based on positional errors of supernova images, finding a slightly tighter constraint of $H_0=70.3\pm4.4 \Mpc$. Compared to the constraint on $H_0=64.8^{+4.4}_{-4.3}\Mpc$ derived from eight cluster lens models in Ref.~\cite{2023Sci...380.1322K}, our results represent a comparable precision, although the central value is about $1\sigma$ higher in our result.

Our finding appears to differ from our previous work, which revealed significant lens mass model dependence in the cluster-lensed quasar system SDSS J1004+4112 and yielded a weak constraint on the Hubble constant of $H_0=67.5^{+14.5}_{-8.9}\Mpc$ \cite{2023PhRvD.108h3532L}. We argue that the much larger number of multiple images for SN Refsdal than for SDSS J1004+4112 plays a crucial role in breaking the degeneracy among different lens mass models and in obtaining the more accurate measurement of the Hubble constant.

In addition, we examine critical curves, the radial convergence, the tangential shear, and magnification factors of the multiply lensed supernova images, and their correlations with the best-fitting Hubble constant values. 
We observe correlations between the Hubble constant and the radial convergence and tangential shear profiles as well as magnification factors. While such correlation has been known for galaxy-scale lenses, our analysis well demonstrates that such correlation exists also for the analysis of massive clusters with complex mass distributions and many multiple images. Furthermore, our analysis clearly demonstrates the usefulness of gravitationally lensed Type Ia supernovae, even for the cluster-scale.

The correlation between radial density profiles and  best-fitting Hubble constant values indicates that future observations conducted with advanced telescopes and instrumentation, such as the James Webb Space Telescope, 
can improve constraint on the Hubble constant from SN Refsdal, by finding many more multiple images.

\begin{acknowledgments}
We thank anonymous referees for useful comments and suggestions.
YL was supported by the China Scholarship Council (Grant No. 202106040084). 
MO was supported by JSPS KAKENHI Grant Numbers JP25H00662, JP25H00672, and JP22K21349.
\end{acknowledgments}



\bibliography{refer}

\begin{thebibliography}{71}
\expandafter\ifx\csname natexlab\endcsname\relax\def\natexlab#1{#1}\fi
\expandafter\ifx\csname bibnamefont\endcsname\relax
  \def\bibnamefont#1{#1}\fi
\expandafter\ifx\csname bibfnamefont\endcsname\relax
  \def\bibfnamefont#1{#1}\fi
\expandafter\ifx\csname citenamefont\endcsname\relax
  \def\citenamefont#1{#1}\fi
\expandafter\ifx\csname url\endcsname\relax
  \def\url#1{\texttt{#1}}\fi
\expandafter\ifx\csname urlprefix\endcsname\relax\def\urlprefix{URL }\fi
\providecommand{\bibinfo}[2]{#2}
\providecommand{\eprint}[2][]{\url{#2}}

\bibitem[{\citenamefont{{Hubble}}(1929)}]{1929PNAS...15..168H}
\bibinfo{author}{\bibfnamefont{E.}~\bibnamefont{{Hubble}}},
  \bibinfo{journal}{Proceedings of the National Academy of Science}
  \textbf{\bibinfo{volume}{15}}, \bibinfo{pages}{168} (\bibinfo{year}{1929}).

\bibitem[{\citenamefont{{Lema{\^\i}tre}}(1931)}]{1931MNRAS..91..483L}
\bibinfo{author}{\bibfnamefont{G.}~\bibnamefont{{Lema{\^\i}tre}}},
  \bibinfo{journal}{\mnras} \textbf{\bibinfo{volume}{91}}, \bibinfo{pages}{483}
  (\bibinfo{year}{1931}).

\bibitem[{\citenamefont{{Freedman} et~al.}(2001)\citenamefont{{Freedman},
  {Madore}, {Gibson}, {Ferrarese}, {Kelson}, {Sakai}, {Mould}, {Kennicutt},
  {Ford}, {Graham} et~al.}}]{2001ApJ...553...47F}
\bibinfo{author}{\bibfnamefont{W.~L.} \bibnamefont{{Freedman}}},
  \bibinfo{author}{\bibfnamefont{B.~F.} \bibnamefont{{Madore}}},
  \bibinfo{author}{\bibfnamefont{B.~K.} \bibnamefont{{Gibson}}},
  \bibinfo{author}{\bibfnamefont{L.}~\bibnamefont{{Ferrarese}}},
  \bibinfo{author}{\bibfnamefont{D.~D.} \bibnamefont{{Kelson}}},
  \bibinfo{author}{\bibfnamefont{S.}~\bibnamefont{{Sakai}}},
  \bibinfo{author}{\bibfnamefont{J.~R.} \bibnamefont{{Mould}}},
  \bibinfo{author}{\bibfnamefont{J.}~\bibnamefont{{Kennicutt}},
  \bibfnamefont{Robert~C.}}, \bibinfo{author}{\bibfnamefont{H.~C.}
  \bibnamefont{{Ford}}}, \bibinfo{author}{\bibfnamefont{J.~A.}
  \bibnamefont{{Graham}}}, \bibnamefont{et~al.}, \bibinfo{journal}{\apj}
  \textbf{\bibinfo{volume}{553}}, \bibinfo{pages}{47} (\bibinfo{year}{2001}),
  \eprint{astro-ph/0012376}.

\bibitem[{\citenamefont{{Nussbaumer} and {Bieri}}(2011)}]{2011Obs...131..394N}
\bibinfo{author}{\bibfnamefont{H.}~\bibnamefont{{Nussbaumer}}}
  \bibnamefont{and} \bibinfo{author}{\bibfnamefont{L.}~\bibnamefont{{Bieri}}},
  \bibinfo{journal}{The Observatory} \textbf{\bibinfo{volume}{131}},
  \bibinfo{pages}{394} (\bibinfo{year}{2011}), \eprint{1107.2281}.

\bibitem[{\citenamefont{{Weinberg} et~al.}(2013)\citenamefont{{Weinberg},
  {Mortonson}, {Eisenstein}, {Hirata}, {Riess}, and
  {Rozo}}}]{2013PhR...530...87W}
\bibinfo{author}{\bibfnamefont{D.~H.} \bibnamefont{{Weinberg}}},
  \bibinfo{author}{\bibfnamefont{M.~J.} \bibnamefont{{Mortonson}}},
  \bibinfo{author}{\bibfnamefont{D.~J.} \bibnamefont{{Eisenstein}}},
  \bibinfo{author}{\bibfnamefont{C.}~\bibnamefont{{Hirata}}},
  \bibinfo{author}{\bibfnamefont{A.~G.} \bibnamefont{{Riess}}},
  \bibnamefont{and} \bibinfo{author}{\bibfnamefont{E.}~\bibnamefont{{Rozo}}},
  \bibinfo{journal}{\physrep} \textbf{\bibinfo{volume}{530}},
  \bibinfo{pages}{87} (\bibinfo{year}{2013}), \eprint{1201.2434}.

\bibitem[{\citenamefont{{Verde} et~al.}(2019)\citenamefont{{Verde}, {Treu}, and
  {Riess}}}]{2019NatAs...3..891V}
\bibinfo{author}{\bibfnamefont{L.}~\bibnamefont{{Verde}}},
  \bibinfo{author}{\bibfnamefont{T.}~\bibnamefont{{Treu}}}, \bibnamefont{and}
  \bibinfo{author}{\bibfnamefont{A.~G.} \bibnamefont{{Riess}}},
  \bibinfo{journal}{Nature Astronomy} \textbf{\bibinfo{volume}{3}},
  \bibinfo{pages}{891} (\bibinfo{year}{2019}), \eprint{1907.10625}.

\bibitem[{\citenamefont{{Freedman}}(2021)}]{2021ApJ...919...16F}
\bibinfo{author}{\bibfnamefont{W.~L.} \bibnamefont{{Freedman}}},
  \bibinfo{journal}{\apj} \textbf{\bibinfo{volume}{919}}, \bibinfo{eid}{16}
  (\bibinfo{year}{2021}), \eprint{2106.15656}.

\bibitem[{\citenamefont{{Abdalla} et~al.}(2022)\citenamefont{{Abdalla},
  {Abell{\'a}n}, {Aboubrahim}, {Agnello}, {Akarsu}, {Akrami}, {Alestas},
  {Aloni}, {Amendola}, {Anchordoqui} et~al.}}]{2022JHEAp..34...49A}
\bibinfo{author}{\bibfnamefont{E.}~\bibnamefont{{Abdalla}}},
  \bibinfo{author}{\bibfnamefont{G.~F.} \bibnamefont{{Abell{\'a}n}}},
  \bibinfo{author}{\bibfnamefont{A.}~\bibnamefont{{Aboubrahim}}},
  \bibinfo{author}{\bibfnamefont{A.}~\bibnamefont{{Agnello}}},
  \bibinfo{author}{\bibfnamefont{{\"O}.}~\bibnamefont{{Akarsu}}},
  \bibinfo{author}{\bibfnamefont{Y.}~\bibnamefont{{Akrami}}},
  \bibinfo{author}{\bibfnamefont{G.}~\bibnamefont{{Alestas}}},
  \bibinfo{author}{\bibfnamefont{D.}~\bibnamefont{{Aloni}}},
  \bibinfo{author}{\bibfnamefont{L.}~\bibnamefont{{Amendola}}},
  \bibinfo{author}{\bibfnamefont{L.~A.} \bibnamefont{{Anchordoqui}}},
  \bibnamefont{et~al.}, \bibinfo{journal}{Journal of High Energy Astrophysics}
  \textbf{\bibinfo{volume}{34}}, \bibinfo{pages}{49} (\bibinfo{year}{2022}),
  \eprint{2203.06142}.

\bibitem[{\citenamefont{{Hu} and {Wang}}(2023)}]{2023Univ....9...94H}
\bibinfo{author}{\bibfnamefont{J.-P.} \bibnamefont{{Hu}}} \bibnamefont{and}
  \bibinfo{author}{\bibfnamefont{F.-Y.} \bibnamefont{{Wang}}},
  \bibinfo{journal}{Universe} \textbf{\bibinfo{volume}{9}}, \bibinfo{pages}{94}
  (\bibinfo{year}{2023}), \eprint{2302.05709}.

\bibitem[{\citenamefont{{Linder}}(2023)}]{2023arXiv230109695L}
\bibinfo{author}{\bibfnamefont{E.~V.} \bibnamefont{{Linder}}},
  \bibinfo{journal}{arXiv e-prints} \bibinfo{eid}{arXiv:2301.09695}
  (\bibinfo{year}{2023}), \eprint{2301.09695}.

\bibitem[{\citenamefont{{Cervantes-Cota}
  et~al.}(2023)\citenamefont{{Cervantes-Cota}, {Galindo-Uribarri}, and
  {Smoot}}}]{2023arXiv231107552C}
\bibinfo{author}{\bibfnamefont{J.~L.} \bibnamefont{{Cervantes-Cota}}},
  \bibinfo{author}{\bibfnamefont{S.}~\bibnamefont{{Galindo-Uribarri}}},
  \bibnamefont{and} \bibinfo{author}{\bibfnamefont{G.~F.}
  \bibnamefont{{Smoot}}}, \bibinfo{journal}{arXiv e-prints}
  \bibinfo{eid}{arXiv:2311.07552} (\bibinfo{year}{2023}), \eprint{2311.07552}.

\bibitem[{\citenamefont{{Riess} et~al.}(2022)\citenamefont{{Riess}, {Yuan},
  {Macri}, {Scolnic}, {Brout}, {Casertano}, {Jones}, {Murakami}, {Anand},
  {Breuval} et~al.}}]{2022ApJ...934L...7R}
\bibinfo{author}{\bibfnamefont{A.~G.} \bibnamefont{{Riess}}},
  \bibinfo{author}{\bibfnamefont{W.}~\bibnamefont{{Yuan}}},
  \bibinfo{author}{\bibfnamefont{L.~M.} \bibnamefont{{Macri}}},
  \bibinfo{author}{\bibfnamefont{D.}~\bibnamefont{{Scolnic}}},
  \bibinfo{author}{\bibfnamefont{D.}~\bibnamefont{{Brout}}},
  \bibinfo{author}{\bibfnamefont{S.}~\bibnamefont{{Casertano}}},
  \bibinfo{author}{\bibfnamefont{D.~O.} \bibnamefont{{Jones}}},
  \bibinfo{author}{\bibfnamefont{Y.}~\bibnamefont{{Murakami}}},
  \bibinfo{author}{\bibfnamefont{G.~S.} \bibnamefont{{Anand}}},
  \bibinfo{author}{\bibfnamefont{L.}~\bibnamefont{{Breuval}}},
  \bibnamefont{et~al.}, \bibinfo{journal}{\apjl}
  \textbf{\bibinfo{volume}{934}}, \bibinfo{eid}{L7} (\bibinfo{year}{2022}),
  \eprint{2112.04510}.

\bibitem[{\citenamefont{{Planck Collaboration}
  et~al.}(2020)\citenamefont{{Planck Collaboration}, {Aghanim}, {Akrami},
  {Ashdown}, {Aumont}, {Baccigalupi}, {Ballardini}, {Banday}, {Barreiro},
  {Bartolo} et~al.}}]{2020A&A...641A...6P}
\bibinfo{author}{\bibnamefont{{Planck Collaboration}}},
  \bibinfo{author}{\bibfnamefont{N.}~\bibnamefont{{Aghanim}}},
  \bibinfo{author}{\bibfnamefont{Y.}~\bibnamefont{{Akrami}}},
  \bibinfo{author}{\bibfnamefont{M.}~\bibnamefont{{Ashdown}}},
  \bibinfo{author}{\bibfnamefont{J.}~\bibnamefont{{Aumont}}},
  \bibinfo{author}{\bibfnamefont{C.}~\bibnamefont{{Baccigalupi}}},
  \bibinfo{author}{\bibfnamefont{M.}~\bibnamefont{{Ballardini}}},
  \bibinfo{author}{\bibfnamefont{A.~J.} \bibnamefont{{Banday}}},
  \bibinfo{author}{\bibfnamefont{R.~B.} \bibnamefont{{Barreiro}}},
  \bibinfo{author}{\bibfnamefont{N.}~\bibnamefont{{Bartolo}}},
  \bibnamefont{et~al.}, \bibinfo{journal}{\aap} \textbf{\bibinfo{volume}{641}},
  \bibinfo{eid}{A6} (\bibinfo{year}{2020}), \eprint{1807.06209}.

\bibitem[{\citenamefont{{Dietrich} et~al.}(2020)\citenamefont{{Dietrich},
  {Coughlin}, {Pang}, {Bulla}, {Heinzel}, {Issa}, {Tews}, and
  {Antier}}}]{2020Sci...370.1450D}
\bibinfo{author}{\bibfnamefont{T.}~\bibnamefont{{Dietrich}}},
  \bibinfo{author}{\bibfnamefont{M.~W.} \bibnamefont{{Coughlin}}},
  \bibinfo{author}{\bibfnamefont{P.~T.~H.} \bibnamefont{{Pang}}},
  \bibinfo{author}{\bibfnamefont{M.}~\bibnamefont{{Bulla}}},
  \bibinfo{author}{\bibfnamefont{J.}~\bibnamefont{{Heinzel}}},
  \bibinfo{author}{\bibfnamefont{L.}~\bibnamefont{{Issa}}},
  \bibinfo{author}{\bibfnamefont{I.}~\bibnamefont{{Tews}}}, \bibnamefont{and}
  \bibinfo{author}{\bibfnamefont{S.}~\bibnamefont{{Antier}}},
  \bibinfo{journal}{Science} \textbf{\bibinfo{volume}{370}},
  \bibinfo{pages}{1450} (\bibinfo{year}{2020}), \eprint{2002.11355}.

\bibitem[{\citenamefont{{Alfradique} et~al.}(2024)\citenamefont{{Alfradique},
  {Bom}, {Palmese}, {Teixeira}, {Santana-Silva}, {Drlica-Wagner}, {Riley},
  {Mart{\'\i}nez-V{\'a}zquez}, {Sand}, {Stringfellow}
  et~al.}}]{2024MNRAS.tmp...74A}
\bibinfo{author}{\bibfnamefont{V.}~\bibnamefont{{Alfradique}}},
  \bibinfo{author}{\bibfnamefont{C.~R.} \bibnamefont{{Bom}}},
  \bibinfo{author}{\bibfnamefont{A.}~\bibnamefont{{Palmese}}},
  \bibinfo{author}{\bibfnamefont{G.}~\bibnamefont{{Teixeira}}},
  \bibinfo{author}{\bibfnamefont{L.}~\bibnamefont{{Santana-Silva}}},
  \bibinfo{author}{\bibfnamefont{A.}~\bibnamefont{{Drlica-Wagner}}},
  \bibinfo{author}{\bibfnamefont{A.~H.} \bibnamefont{{Riley}}},
  \bibinfo{author}{\bibfnamefont{C.~E.}
  \bibnamefont{{Mart{\'\i}nez-V{\'a}zquez}}},
  \bibinfo{author}{\bibfnamefont{D.~J.} \bibnamefont{{Sand}}},
  \bibinfo{author}{\bibfnamefont{G.~S.} \bibnamefont{{Stringfellow}}},
  \bibnamefont{et~al.}, \bibinfo{journal}{\mnras}  (\bibinfo{year}{2024}),
  \eprint{2310.13695}.

\bibitem[{\citenamefont{{James}
  et~al.}(2022{\natexlab{a}})\citenamefont{{James}, {Prochaska}, {Macquart},
  {North-Hickey}, {Bannister}, and {Dunning}}}]{2022MNRAS.509.4775J}
\bibinfo{author}{\bibfnamefont{C.~W.} \bibnamefont{{James}}},
  \bibinfo{author}{\bibfnamefont{J.~X.} \bibnamefont{{Prochaska}}},
  \bibinfo{author}{\bibfnamefont{J.~P.} \bibnamefont{{Macquart}}},
  \bibinfo{author}{\bibfnamefont{F.~O.} \bibnamefont{{North-Hickey}}},
  \bibinfo{author}{\bibfnamefont{K.~W.} \bibnamefont{{Bannister}}},
  \bibnamefont{and}
  \bibinfo{author}{\bibfnamefont{A.}~\bibnamefont{{Dunning}}},
  \bibinfo{journal}{\mnras} \textbf{\bibinfo{volume}{509}},
  \bibinfo{pages}{4775} (\bibinfo{year}{2022}{\natexlab{a}}),
  \eprint{2101.08005}.

\bibitem[{\citenamefont{{Hagstotz} et~al.}(2022)\citenamefont{{Hagstotz},
  {Reischke}, and {Lilow}}}]{2022MNRAS.511..662H}
\bibinfo{author}{\bibfnamefont{S.}~\bibnamefont{{Hagstotz}}},
  \bibinfo{author}{\bibfnamefont{R.}~\bibnamefont{{Reischke}}},
  \bibnamefont{and} \bibinfo{author}{\bibfnamefont{R.}~\bibnamefont{{Lilow}}},
  \bibinfo{journal}{\mnras} \textbf{\bibinfo{volume}{511}},
  \bibinfo{pages}{662} (\bibinfo{year}{2022}), \eprint{2104.04538}.

\bibitem[{\citenamefont{{Wu} et~al.}(2022)\citenamefont{{Wu}, {Zhang}, and
  {Wang}}}]{2022MNRAS.515L...1W}
\bibinfo{author}{\bibfnamefont{Q.}~\bibnamefont{{Wu}}},
  \bibinfo{author}{\bibfnamefont{G.-Q.} \bibnamefont{{Zhang}}},
  \bibnamefont{and} \bibinfo{author}{\bibfnamefont{F.-Y.}
  \bibnamefont{{Wang}}}, \bibinfo{journal}{\mnras}
  \textbf{\bibinfo{volume}{515}}, \bibinfo{pages}{L1} (\bibinfo{year}{2022}),
  \eprint{2108.00581}.

\bibitem[{\citenamefont{{James}
  et~al.}(2022{\natexlab{b}})\citenamefont{{James}, {Ghosh}, {Prochaska},
  {Bannister}, {Bhandari}, {Day}, {Deller}, {Glowacki}, {Gordon}, {Heintz}
  et~al.}}]{2022MNRAS.516.4862J}
\bibinfo{author}{\bibfnamefont{C.~W.} \bibnamefont{{James}}},
  \bibinfo{author}{\bibfnamefont{E.~M.} \bibnamefont{{Ghosh}}},
  \bibinfo{author}{\bibfnamefont{J.~X.} \bibnamefont{{Prochaska}}},
  \bibinfo{author}{\bibfnamefont{K.~W.} \bibnamefont{{Bannister}}},
  \bibinfo{author}{\bibfnamefont{S.}~\bibnamefont{{Bhandari}}},
  \bibinfo{author}{\bibfnamefont{C.~K.} \bibnamefont{{Day}}},
  \bibinfo{author}{\bibfnamefont{A.~T.} \bibnamefont{{Deller}}},
  \bibinfo{author}{\bibfnamefont{M.}~\bibnamefont{{Glowacki}}},
  \bibinfo{author}{\bibfnamefont{A.~C.} \bibnamefont{{Gordon}}},
  \bibinfo{author}{\bibfnamefont{K.~E.} \bibnamefont{{Heintz}}},
  \bibnamefont{et~al.}, \bibinfo{journal}{\mnras}
  \textbf{\bibinfo{volume}{516}}, \bibinfo{pages}{4862}
  (\bibinfo{year}{2022}{\natexlab{b}}), \eprint{2208.00819}.

\bibitem[{\citenamefont{{Refsdal}}(1964)}]{1964MNRAS.128..307R}
\bibinfo{author}{\bibfnamefont{S.}~\bibnamefont{{Refsdal}}},
  \bibinfo{journal}{\mnras} \textbf{\bibinfo{volume}{128}},
  \bibinfo{pages}{307} (\bibinfo{year}{1964}).

\bibitem[{\citenamefont{{Jee} et~al.}(2019)\citenamefont{{Jee}, {Suyu},
  {Komatsu}, {Fassnacht}, {Hilbert}, and {Koopmans}}}]{2019Sci...365.1134J}
\bibinfo{author}{\bibfnamefont{I.}~\bibnamefont{{Jee}}},
  \bibinfo{author}{\bibfnamefont{S.~H.} \bibnamefont{{Suyu}}},
  \bibinfo{author}{\bibfnamefont{E.}~\bibnamefont{{Komatsu}}},
  \bibinfo{author}{\bibfnamefont{C.~D.} \bibnamefont{{Fassnacht}}},
  \bibinfo{author}{\bibfnamefont{S.}~\bibnamefont{{Hilbert}}},
  \bibnamefont{and} \bibinfo{author}{\bibfnamefont{L.~V.~E.}
  \bibnamefont{{Koopmans}}}, \bibinfo{journal}{Science}
  \textbf{\bibinfo{volume}{365}}, \bibinfo{pages}{1134} (\bibinfo{year}{2019}),
  \eprint{1909.06712}.

\bibitem[{\citenamefont{{Chen} et~al.}(2019)\citenamefont{{Chen}, {Fassnacht},
  {Suyu}, {Rusu}, {Chan}, {Wong}, {Auger}, {Hilbert}, {Bonvin}, {Birrer}
  et~al.}}]{2019MNRAS.490.1743C}
\bibinfo{author}{\bibfnamefont{G.~C.~F.} \bibnamefont{{Chen}}},
  \bibinfo{author}{\bibfnamefont{C.~D.} \bibnamefont{{Fassnacht}}},
  \bibinfo{author}{\bibfnamefont{S.~H.} \bibnamefont{{Suyu}}},
  \bibinfo{author}{\bibfnamefont{C.~E.} \bibnamefont{{Rusu}}},
  \bibinfo{author}{\bibfnamefont{J.~H.~H.} \bibnamefont{{Chan}}},
  \bibinfo{author}{\bibfnamefont{K.~C.} \bibnamefont{{Wong}}},
  \bibinfo{author}{\bibfnamefont{M.~W.} \bibnamefont{{Auger}}},
  \bibinfo{author}{\bibfnamefont{S.}~\bibnamefont{{Hilbert}}},
  \bibinfo{author}{\bibfnamefont{V.}~\bibnamefont{{Bonvin}}},
  \bibinfo{author}{\bibfnamefont{S.}~\bibnamefont{{Birrer}}},
  \bibnamefont{et~al.}, \bibinfo{journal}{\mnras}
  \textbf{\bibinfo{volume}{490}}, \bibinfo{pages}{1743} (\bibinfo{year}{2019}),
  \eprint{1907.02533}.

\bibitem[{\citenamefont{{Wong} et~al.}(2017)\citenamefont{{Wong}, {Suyu},
  {Auger}, {Bonvin}, {Courbin}, {Fassnacht}, {Halkola}, {Rusu}, {Sluse},
  {Sonnenfeld} et~al.}}]{2017MNRAS.465.4895W}
\bibinfo{author}{\bibfnamefont{K.~C.} \bibnamefont{{Wong}}},
  \bibinfo{author}{\bibfnamefont{S.~H.} \bibnamefont{{Suyu}}},
  \bibinfo{author}{\bibfnamefont{M.~W.} \bibnamefont{{Auger}}},
  \bibinfo{author}{\bibfnamefont{V.}~\bibnamefont{{Bonvin}}},
  \bibinfo{author}{\bibfnamefont{F.}~\bibnamefont{{Courbin}}},
  \bibinfo{author}{\bibfnamefont{C.~D.} \bibnamefont{{Fassnacht}}},
  \bibinfo{author}{\bibfnamefont{A.}~\bibnamefont{{Halkola}}},
  \bibinfo{author}{\bibfnamefont{C.~E.} \bibnamefont{{Rusu}}},
  \bibinfo{author}{\bibfnamefont{D.}~\bibnamefont{{Sluse}}},
  \bibinfo{author}{\bibfnamefont{A.}~\bibnamefont{{Sonnenfeld}}},
  \bibnamefont{et~al.}, \bibinfo{journal}{\mnras}
  \textbf{\bibinfo{volume}{465}}, \bibinfo{pages}{4895} (\bibinfo{year}{2017}),
  \eprint{1607.01403}.

\bibitem[{\citenamefont{{Birrer} et~al.}(2019)\citenamefont{{Birrer}, {Treu},
  {Rusu}, {Bonvin}, {Fassnacht}, {Chan}, {Agnello}, {Shajib}, {Chen}, {Auger}
  et~al.}}]{2019MNRAS.484.4726B}
\bibinfo{author}{\bibfnamefont{S.}~\bibnamefont{{Birrer}}},
  \bibinfo{author}{\bibfnamefont{T.}~\bibnamefont{{Treu}}},
  \bibinfo{author}{\bibfnamefont{C.~E.} \bibnamefont{{Rusu}}},
  \bibinfo{author}{\bibfnamefont{V.}~\bibnamefont{{Bonvin}}},
  \bibinfo{author}{\bibfnamefont{C.~D.} \bibnamefont{{Fassnacht}}},
  \bibinfo{author}{\bibfnamefont{J.~H.~H.} \bibnamefont{{Chan}}},
  \bibinfo{author}{\bibfnamefont{A.}~\bibnamefont{{Agnello}}},
  \bibinfo{author}{\bibfnamefont{A.~J.} \bibnamefont{{Shajib}}},
  \bibinfo{author}{\bibfnamefont{G.~C.~F.} \bibnamefont{{Chen}}},
  \bibinfo{author}{\bibfnamefont{M.}~\bibnamefont{{Auger}}},
  \bibnamefont{et~al.}, \bibinfo{journal}{\mnras}
  \textbf{\bibinfo{volume}{484}}, \bibinfo{pages}{4726} (\bibinfo{year}{2019}),
  \eprint{1809.01274}.

\bibitem[{\citenamefont{{Shajib} et~al.}(2020)\citenamefont{{Shajib}, {Birrer},
  {Treu}, {Agnello}, {Buckley-Geer}, {Chan}, {Christensen}, {Lemon}, {Lin},
  {Millon} et~al.}}]{2020MNRAS.494.6072S}
\bibinfo{author}{\bibfnamefont{A.~J.} \bibnamefont{{Shajib}}},
  \bibinfo{author}{\bibfnamefont{S.}~\bibnamefont{{Birrer}}},
  \bibinfo{author}{\bibfnamefont{T.}~\bibnamefont{{Treu}}},
  \bibinfo{author}{\bibfnamefont{A.}~\bibnamefont{{Agnello}}},
  \bibinfo{author}{\bibfnamefont{E.~J.} \bibnamefont{{Buckley-Geer}}},
  \bibinfo{author}{\bibfnamefont{J.~H.~H.} \bibnamefont{{Chan}}},
  \bibinfo{author}{\bibfnamefont{L.}~\bibnamefont{{Christensen}}},
  \bibinfo{author}{\bibfnamefont{C.}~\bibnamefont{{Lemon}}},
  \bibinfo{author}{\bibfnamefont{H.}~\bibnamefont{{Lin}}},
  \bibinfo{author}{\bibfnamefont{M.}~\bibnamefont{{Millon}}},
  \bibnamefont{et~al.}, \bibinfo{journal}{\mnras}
  \textbf{\bibinfo{volume}{494}}, \bibinfo{pages}{6072} (\bibinfo{year}{2020}),
  \eprint{1910.06306}.

\bibitem[{\citenamefont{{Lin} et~al.}(2017)\citenamefont{{Lin}, {Buckley-Geer},
  {Agnello}, {Ostrovski}, {McMahon}, {Nord}, {Kuropatkin}, {Tucker}, {Treu},
  {Chan} et~al.}}]{2017ApJ...838L..15L}
\bibinfo{author}{\bibfnamefont{H.}~\bibnamefont{{Lin}}},
  \bibinfo{author}{\bibfnamefont{E.}~\bibnamefont{{Buckley-Geer}}},
  \bibinfo{author}{\bibfnamefont{A.}~\bibnamefont{{Agnello}}},
  \bibinfo{author}{\bibfnamefont{F.}~\bibnamefont{{Ostrovski}}},
  \bibinfo{author}{\bibfnamefont{R.~G.} \bibnamefont{{McMahon}}},
  \bibinfo{author}{\bibfnamefont{B.}~\bibnamefont{{Nord}}},
  \bibinfo{author}{\bibfnamefont{N.}~\bibnamefont{{Kuropatkin}}},
  \bibinfo{author}{\bibfnamefont{D.~L.} \bibnamefont{{Tucker}}},
  \bibinfo{author}{\bibfnamefont{T.}~\bibnamefont{{Treu}}},
  \bibinfo{author}{\bibfnamefont{J.~H.~H.} \bibnamefont{{Chan}}},
  \bibnamefont{et~al.}, \bibinfo{journal}{\apjl}
  \textbf{\bibinfo{volume}{838}}, \bibinfo{eid}{L15} (\bibinfo{year}{2017}),
  \eprint{1702.00072}.

\bibitem[{\citenamefont{{Shajib} et~al.}(2023)\citenamefont{{Shajib},
  {Mozumdar}, {Chen}, {Treu}, {Cappellari}, {Knabel}, {Suyu}, {Bennert},
  {Frieman}, {Sluse} et~al.}}]{2023A&A...673A...9S}
\bibinfo{author}{\bibfnamefont{A.~J.} \bibnamefont{{Shajib}}},
  \bibinfo{author}{\bibfnamefont{P.}~\bibnamefont{{Mozumdar}}},
  \bibinfo{author}{\bibfnamefont{G.~C.~F.} \bibnamefont{{Chen}}},
  \bibinfo{author}{\bibfnamefont{T.}~\bibnamefont{{Treu}}},
  \bibinfo{author}{\bibfnamefont{M.}~\bibnamefont{{Cappellari}}},
  \bibinfo{author}{\bibfnamefont{S.}~\bibnamefont{{Knabel}}},
  \bibinfo{author}{\bibfnamefont{S.~H.} \bibnamefont{{Suyu}}},
  \bibinfo{author}{\bibfnamefont{V.~N.} \bibnamefont{{Bennert}}},
  \bibinfo{author}{\bibfnamefont{J.~A.} \bibnamefont{{Frieman}}},
  \bibinfo{author}{\bibfnamefont{D.}~\bibnamefont{{Sluse}}},
  \bibnamefont{et~al.}, \bibinfo{journal}{\aap} \textbf{\bibinfo{volume}{673}},
  \bibinfo{eid}{A9} (\bibinfo{year}{2023}), \eprint{2301.02656}.

\bibitem[{\citenamefont{{Denzel} et~al.}(2021)\citenamefont{{Denzel}, {Coles},
  {Saha}, and {Williams}}}]{2021MNRAS.501..784D}
\bibinfo{author}{\bibfnamefont{P.}~\bibnamefont{{Denzel}}},
  \bibinfo{author}{\bibfnamefont{J.~P.} \bibnamefont{{Coles}}},
  \bibinfo{author}{\bibfnamefont{P.}~\bibnamefont{{Saha}}}, \bibnamefont{and}
  \bibinfo{author}{\bibfnamefont{L.~L.~R.} \bibnamefont{{Williams}}},
  \bibinfo{journal}{\mnras} \textbf{\bibinfo{volume}{501}},
  \bibinfo{pages}{784} (\bibinfo{year}{2021}), \eprint{2007.14398}.

\bibitem[{\citenamefont{{Napier} et~al.}(2023)\citenamefont{{Napier}, {Sharon},
  {Dahle}, {Bayliss}, {Gladders}, {Mahler}, {Rigby}, and
  {Florian}}}]{2023arXiv230111240N}
\bibinfo{author}{\bibfnamefont{K.}~\bibnamefont{{Napier}}},
  \bibinfo{author}{\bibfnamefont{K.}~\bibnamefont{{Sharon}}},
  \bibinfo{author}{\bibfnamefont{H.}~\bibnamefont{{Dahle}}},
  \bibinfo{author}{\bibfnamefont{M.}~\bibnamefont{{Bayliss}}},
  \bibinfo{author}{\bibfnamefont{M.~D.} \bibnamefont{{Gladders}}},
  \bibinfo{author}{\bibfnamefont{G.}~\bibnamefont{{Mahler}}},
  \bibinfo{author}{\bibfnamefont{J.~R.} \bibnamefont{{Rigby}}},
  \bibnamefont{and}
  \bibinfo{author}{\bibfnamefont{M.}~\bibnamefont{{Florian}}},
  \bibinfo{journal}{arXiv e-prints} \bibinfo{eid}{arXiv:2301.11240}
  (\bibinfo{year}{2023}), \eprint{2301.11240}.

\bibitem[{\citenamefont{{Mart{\'\i}nez-Arrizabalaga}
  et~al.}(2023)\citenamefont{{Mart{\'\i}nez-Arrizabalaga}, {Diego}, and
  {Goicoechea}}}]{2023arXiv230914776M}
\bibinfo{author}{\bibfnamefont{J.}~\bibnamefont{{Mart{\'\i}nez-Arrizabalaga}}},
  \bibinfo{author}{\bibfnamefont{J.~M.} \bibnamefont{{Diego}}},
  \bibnamefont{and} \bibinfo{author}{\bibfnamefont{L.~J.}
  \bibnamefont{{Goicoechea}}}, \bibinfo{journal}{arXiv e-prints}
  \bibinfo{eid}{arXiv:2309.14776} (\bibinfo{year}{2023}), \eprint{2309.14776}.

\bibitem[{\citenamefont{{Meneghetti} et~al.}(2017)\citenamefont{{Meneghetti},
  {Natarajan}, {Coe}, {Contini}, {De Lucia}, {Giocoli}, {Acebron}, {Borgani},
  {Bradac}, {Diego} et~al.}}]{2017MNRAS.472.3177M}
\bibinfo{author}{\bibfnamefont{M.}~\bibnamefont{{Meneghetti}}},
  \bibinfo{author}{\bibfnamefont{P.}~\bibnamefont{{Natarajan}}},
  \bibinfo{author}{\bibfnamefont{D.}~\bibnamefont{{Coe}}},
  \bibinfo{author}{\bibfnamefont{E.}~\bibnamefont{{Contini}}},
  \bibinfo{author}{\bibfnamefont{G.}~\bibnamefont{{De Lucia}}},
  \bibinfo{author}{\bibfnamefont{C.}~\bibnamefont{{Giocoli}}},
  \bibinfo{author}{\bibfnamefont{A.}~\bibnamefont{{Acebron}}},
  \bibinfo{author}{\bibfnamefont{S.}~\bibnamefont{{Borgani}}},
  \bibinfo{author}{\bibfnamefont{M.}~\bibnamefont{{Bradac}}},
  \bibinfo{author}{\bibfnamefont{J.~M.} \bibnamefont{{Diego}}},
  \bibnamefont{et~al.}, \bibinfo{journal}{\mnras}
  \textbf{\bibinfo{volume}{472}}, \bibinfo{pages}{3177} (\bibinfo{year}{2017}),
  \eprint{1606.04548}.

\bibitem[{\citenamefont{{Liu} et~al.}(2023)\citenamefont{{Liu}, {Oguri}, and
  {Cao}}}]{2023PhRvD.108h3532L}
\bibinfo{author}{\bibfnamefont{Y.}~\bibnamefont{{Liu}}},
  \bibinfo{author}{\bibfnamefont{M.}~\bibnamefont{{Oguri}}}, \bibnamefont{and}
  \bibinfo{author}{\bibfnamefont{S.}~\bibnamefont{{Cao}}},
  \bibinfo{journal}{\prd} \textbf{\bibinfo{volume}{108}}, \bibinfo{eid}{083532}
  (\bibinfo{year}{2023}), \eprint{2307.14833}.

\bibitem[{\citenamefont{{Kelly} et~al.}(2015)\citenamefont{{Kelly}, {Rodney},
  {Treu}, {Foley}, {Brammer}, {Schmidt}, {Zitrin}, {Sonnenfeld}, {Strolger},
  {Graur} et~al.}}]{2015Sci...347.1123K}
\bibinfo{author}{\bibfnamefont{P.~L.} \bibnamefont{{Kelly}}},
  \bibinfo{author}{\bibfnamefont{S.~A.} \bibnamefont{{Rodney}}},
  \bibinfo{author}{\bibfnamefont{T.}~\bibnamefont{{Treu}}},
  \bibinfo{author}{\bibfnamefont{R.~J.} \bibnamefont{{Foley}}},
  \bibinfo{author}{\bibfnamefont{G.}~\bibnamefont{{Brammer}}},
  \bibinfo{author}{\bibfnamefont{K.~B.} \bibnamefont{{Schmidt}}},
  \bibinfo{author}{\bibfnamefont{A.}~\bibnamefont{{Zitrin}}},
  \bibinfo{author}{\bibfnamefont{A.}~\bibnamefont{{Sonnenfeld}}},
  \bibinfo{author}{\bibfnamefont{L.-G.} \bibnamefont{{Strolger}}},
  \bibinfo{author}{\bibfnamefont{O.}~\bibnamefont{{Graur}}},
  \bibnamefont{et~al.}, \bibinfo{journal}{Science}
  \textbf{\bibinfo{volume}{347}}, \bibinfo{pages}{1123} (\bibinfo{year}{2015}),
  \eprint{1411.6009}.

\bibitem[{\citenamefont{{Kelly} et~al.}(2016)\citenamefont{{Kelly}, {Rodney},
  {Treu}, {Strolger}, {Foley}, {Jha}, {Selsing}, {Brammer}, {Brada{\v{c}}},
  {Cenko} et~al.}}]{2016ApJ...819L...8K}
\bibinfo{author}{\bibfnamefont{P.~L.} \bibnamefont{{Kelly}}},
  \bibinfo{author}{\bibfnamefont{S.~A.} \bibnamefont{{Rodney}}},
  \bibinfo{author}{\bibfnamefont{T.}~\bibnamefont{{Treu}}},
  \bibinfo{author}{\bibfnamefont{L.~G.} \bibnamefont{{Strolger}}},
  \bibinfo{author}{\bibfnamefont{R.~J.} \bibnamefont{{Foley}}},
  \bibinfo{author}{\bibfnamefont{S.~W.} \bibnamefont{{Jha}}},
  \bibinfo{author}{\bibfnamefont{J.}~\bibnamefont{{Selsing}}},
  \bibinfo{author}{\bibfnamefont{G.}~\bibnamefont{{Brammer}}},
  \bibinfo{author}{\bibfnamefont{M.}~\bibnamefont{{Brada{\v{c}}}}},
  \bibinfo{author}{\bibfnamefont{S.~B.} \bibnamefont{{Cenko}}},
  \bibnamefont{et~al.}, \bibinfo{journal}{\apjl}
  \textbf{\bibinfo{volume}{819}}, \bibinfo{eid}{L8} (\bibinfo{year}{2016}),
  \eprint{1512.04654}.

\bibitem[{\citenamefont{{Vega-Ferrero}
  et~al.}(2018)\citenamefont{{Vega-Ferrero}, {Diego}, {Miranda}, and
  {Bernstein}}}]{2018ApJ...853L..31V}
\bibinfo{author}{\bibfnamefont{J.}~\bibnamefont{{Vega-Ferrero}}},
  \bibinfo{author}{\bibfnamefont{J.~M.} \bibnamefont{{Diego}}},
  \bibinfo{author}{\bibfnamefont{V.}~\bibnamefont{{Miranda}}},
  \bibnamefont{and} \bibinfo{author}{\bibfnamefont{G.~M.}
  \bibnamefont{{Bernstein}}}, \bibinfo{journal}{\apjl}
  \textbf{\bibinfo{volume}{853}}, \bibinfo{eid}{L31} (\bibinfo{year}{2018}),
  \eprint{1712.05800}.

\bibitem[{\citenamefont{{Kelly}
  et~al.}(2023{\natexlab{a}})\citenamefont{{Kelly}, {Rodney}, {Treu}, {Oguri},
  {Chen}, {Zitrin}, {Birrer}, {Bonvin}, {Dessart}, {Diego}
  et~al.}}]{2023Sci...380.1322K}
\bibinfo{author}{\bibfnamefont{P.~L.} \bibnamefont{{Kelly}}},
  \bibinfo{author}{\bibfnamefont{S.}~\bibnamefont{{Rodney}}},
  \bibinfo{author}{\bibfnamefont{T.}~\bibnamefont{{Treu}}},
  \bibinfo{author}{\bibfnamefont{M.}~\bibnamefont{{Oguri}}},
  \bibinfo{author}{\bibfnamefont{W.}~\bibnamefont{{Chen}}},
  \bibinfo{author}{\bibfnamefont{A.}~\bibnamefont{{Zitrin}}},
  \bibinfo{author}{\bibfnamefont{S.}~\bibnamefont{{Birrer}}},
  \bibinfo{author}{\bibfnamefont{V.}~\bibnamefont{{Bonvin}}},
  \bibinfo{author}{\bibfnamefont{L.}~\bibnamefont{{Dessart}}},
  \bibinfo{author}{\bibfnamefont{J.~M.} \bibnamefont{{Diego}}},
  \bibnamefont{et~al.}, \bibinfo{journal}{Science}
  \textbf{\bibinfo{volume}{380}}, \bibinfo{eid}{abh1322}
  (\bibinfo{year}{2023}{\natexlab{a}}), \eprint{2305.06367}.

\bibitem[{\citenamefont{{Grillo} et~al.}(2024)\citenamefont{{Grillo}, {Pagano},
  {Rosati}, and {Suyu}}}]{2024arXiv240110980G}
\bibinfo{author}{\bibfnamefont{C.}~\bibnamefont{{Grillo}}},
  \bibinfo{author}{\bibfnamefont{L.}~\bibnamefont{{Pagano}}},
  \bibinfo{author}{\bibfnamefont{P.}~\bibnamefont{{Rosati}}}, \bibnamefont{and}
  \bibinfo{author}{\bibfnamefont{S.~H.} \bibnamefont{{Suyu}}},
  \bibinfo{journal}{arXiv e-prints} \bibinfo{eid}{arXiv:2401.10980}
  (\bibinfo{year}{2024}), \eprint{2401.10980}.

\bibitem[{\citenamefont{{Grillo} et~al.}(2020)\citenamefont{{Grillo}, {Rosati},
  {Suyu}, {Caminha}, {Mercurio}, and {Halkola}}}]{2020ApJ...898...87G}
\bibinfo{author}{\bibfnamefont{C.}~\bibnamefont{{Grillo}}},
  \bibinfo{author}{\bibfnamefont{P.}~\bibnamefont{{Rosati}}},
  \bibinfo{author}{\bibfnamefont{S.~H.} \bibnamefont{{Suyu}}},
  \bibinfo{author}{\bibfnamefont{G.~B.} \bibnamefont{{Caminha}}},
  \bibinfo{author}{\bibfnamefont{A.}~\bibnamefont{{Mercurio}}},
  \bibnamefont{and}
  \bibinfo{author}{\bibfnamefont{A.}~\bibnamefont{{Halkola}}},
  \bibinfo{journal}{\apj} \textbf{\bibinfo{volume}{898}}, \bibinfo{eid}{87}
  (\bibinfo{year}{2020}), \eprint{2001.02232}.

\bibitem[{\citenamefont{{Kawamata} et~al.}(2016)\citenamefont{{Kawamata},
  {Oguri}, {Ishigaki}, {Shimasaku}, and {Ouchi}}}]{2016ApJ...819..114K}
\bibinfo{author}{\bibfnamefont{R.}~\bibnamefont{{Kawamata}}},
  \bibinfo{author}{\bibfnamefont{M.}~\bibnamefont{{Oguri}}},
  \bibinfo{author}{\bibfnamefont{M.}~\bibnamefont{{Ishigaki}}},
  \bibinfo{author}{\bibfnamefont{K.}~\bibnamefont{{Shimasaku}}},
  \bibnamefont{and} \bibinfo{author}{\bibfnamefont{M.}~\bibnamefont{{Ouchi}}},
  \bibinfo{journal}{\apj} \textbf{\bibinfo{volume}{819}}, \bibinfo{eid}{114}
  (\bibinfo{year}{2016}), \eprint{1510.06400}.

\bibitem[{\citenamefont{{Kelly}
  et~al.}(2023{\natexlab{b}})\citenamefont{{Kelly}, {Rodney}, {Treu}, {Birrer},
  {Bonvin}, {Dessart}, {Foley}, {Filippenko}, {Gilman}, {Jha}
  et~al.}}]{2023ApJ...948...93K}
\bibinfo{author}{\bibfnamefont{P.~L.} \bibnamefont{{Kelly}}},
  \bibinfo{author}{\bibfnamefont{S.}~\bibnamefont{{Rodney}}},
  \bibinfo{author}{\bibfnamefont{T.}~\bibnamefont{{Treu}}},
  \bibinfo{author}{\bibfnamefont{S.}~\bibnamefont{{Birrer}}},
  \bibinfo{author}{\bibfnamefont{V.}~\bibnamefont{{Bonvin}}},
  \bibinfo{author}{\bibfnamefont{L.}~\bibnamefont{{Dessart}}},
  \bibinfo{author}{\bibfnamefont{R.~J.} \bibnamefont{{Foley}}},
  \bibinfo{author}{\bibfnamefont{A.~V.} \bibnamefont{{Filippenko}}},
  \bibinfo{author}{\bibfnamefont{D.}~\bibnamefont{{Gilman}}},
  \bibinfo{author}{\bibfnamefont{S.}~\bibnamefont{{Jha}}},
  \bibnamefont{et~al.}, \bibinfo{journal}{\apj} \textbf{\bibinfo{volume}{948}},
  \bibinfo{eid}{93} (\bibinfo{year}{2023}{\natexlab{b}}), \eprint{2305.06377}.

\bibitem[{\citenamefont{{Oguri}}(2015)}]{2015MNRAS.449L..86O}
\bibinfo{author}{\bibfnamefont{M.}~\bibnamefont{{Oguri}}},
  \bibinfo{journal}{\mnras} \textbf{\bibinfo{volume}{449}},
  \bibinfo{pages}{L86} (\bibinfo{year}{2015}), \eprint{1411.6443}.

\bibitem[{\citenamefont{{Schmidt} et~al.}(2014)\citenamefont{{Schmidt}, {Treu},
  {Brammer}, {Brada{\v{c}}}, {Wang}, {Dijkstra}, {Dressler}, {Fontana},
  {Gavazzi}, {Henry} et~al.}}]{2014ApJ...782L..36S}
\bibinfo{author}{\bibfnamefont{K.~B.} \bibnamefont{{Schmidt}}},
  \bibinfo{author}{\bibfnamefont{T.}~\bibnamefont{{Treu}}},
  \bibinfo{author}{\bibfnamefont{G.~B.} \bibnamefont{{Brammer}}},
  \bibinfo{author}{\bibfnamefont{M.}~\bibnamefont{{Brada{\v{c}}}}},
  \bibinfo{author}{\bibfnamefont{X.}~\bibnamefont{{Wang}}},
  \bibinfo{author}{\bibfnamefont{M.}~\bibnamefont{{Dijkstra}}},
  \bibinfo{author}{\bibfnamefont{A.}~\bibnamefont{{Dressler}}},
  \bibinfo{author}{\bibfnamefont{A.}~\bibnamefont{{Fontana}}},
  \bibinfo{author}{\bibfnamefont{R.}~\bibnamefont{{Gavazzi}}},
  \bibinfo{author}{\bibfnamefont{A.~L.} \bibnamefont{{Henry}}},
  \bibnamefont{et~al.}, \bibinfo{journal}{\apjl}
  \textbf{\bibinfo{volume}{782}}, \bibinfo{eid}{L36} (\bibinfo{year}{2014}),
  \eprint{1401.0532}.

\bibitem[{\citenamefont{{Treu} et~al.}(2015)\citenamefont{{Treu}, {Schmidt},
  {Brammer}, {Vulcani}, {Wang}, {Brada{\v{c}}}, {Dijkstra}, {Dressler},
  {Fontana}, {Gavazzi} et~al.}}]{2015ApJ...812..114T}
\bibinfo{author}{\bibfnamefont{T.}~\bibnamefont{{Treu}}},
  \bibinfo{author}{\bibfnamefont{K.~B.} \bibnamefont{{Schmidt}}},
  \bibinfo{author}{\bibfnamefont{G.~B.} \bibnamefont{{Brammer}}},
  \bibinfo{author}{\bibfnamefont{B.}~\bibnamefont{{Vulcani}}},
  \bibinfo{author}{\bibfnamefont{X.}~\bibnamefont{{Wang}}},
  \bibinfo{author}{\bibfnamefont{M.}~\bibnamefont{{Brada{\v{c}}}}},
  \bibinfo{author}{\bibfnamefont{M.}~\bibnamefont{{Dijkstra}}},
  \bibinfo{author}{\bibfnamefont{A.}~\bibnamefont{{Dressler}}},
  \bibinfo{author}{\bibfnamefont{A.}~\bibnamefont{{Fontana}}},
  \bibinfo{author}{\bibfnamefont{R.}~\bibnamefont{{Gavazzi}}},
  \bibnamefont{et~al.}, \bibinfo{journal}{\apj} \textbf{\bibinfo{volume}{812}},
  \bibinfo{eid}{114} (\bibinfo{year}{2015}), \eprint{1509.00475}.

\bibitem[{\citenamefont{{Sharon} and {Johnson}}(2015)}]{2015ApJ...800L..26S}
\bibinfo{author}{\bibfnamefont{K.}~\bibnamefont{{Sharon}}} \bibnamefont{and}
  \bibinfo{author}{\bibfnamefont{T.~L.} \bibnamefont{{Johnson}}},
  \bibinfo{journal}{\apjl} \textbf{\bibinfo{volume}{800}}, \bibinfo{eid}{L26}
  (\bibinfo{year}{2015}), \eprint{1411.6933}.

\bibitem[{\citenamefont{{Diego} et~al.}(2016)\citenamefont{{Diego},
  {Broadhurst}, {Chen}, {Lim}, {Zitrin}, {Chan}, {Coe}, {Ford}, {Lam}, and
  {Zheng}}}]{2016MNRAS.456..356D}
\bibinfo{author}{\bibfnamefont{J.~M.} \bibnamefont{{Diego}}},
  \bibinfo{author}{\bibfnamefont{T.}~\bibnamefont{{Broadhurst}}},
  \bibinfo{author}{\bibfnamefont{C.}~\bibnamefont{{Chen}}},
  \bibinfo{author}{\bibfnamefont{J.}~\bibnamefont{{Lim}}},
  \bibinfo{author}{\bibfnamefont{A.}~\bibnamefont{{Zitrin}}},
  \bibinfo{author}{\bibfnamefont{B.}~\bibnamefont{{Chan}}},
  \bibinfo{author}{\bibfnamefont{D.}~\bibnamefont{{Coe}}},
  \bibinfo{author}{\bibfnamefont{H.~C.} \bibnamefont{{Ford}}},
  \bibinfo{author}{\bibfnamefont{D.}~\bibnamefont{{Lam}}}, \bibnamefont{and}
  \bibinfo{author}{\bibfnamefont{W.}~\bibnamefont{{Zheng}}},
  \bibinfo{journal}{\mnras} \textbf{\bibinfo{volume}{456}},
  \bibinfo{pages}{356} (\bibinfo{year}{2016}), \eprint{1504.05953}.

\bibitem[{\citenamefont{{Ebeling} et~al.}(2007)\citenamefont{{Ebeling},
  {Barrett}, {Donovan}, {Ma}, {Edge}, and {van
  Speybroeck}}}]{2007ApJ...661L..33E}
\bibinfo{author}{\bibfnamefont{H.}~\bibnamefont{{Ebeling}}},
  \bibinfo{author}{\bibfnamefont{E.}~\bibnamefont{{Barrett}}},
  \bibinfo{author}{\bibfnamefont{D.}~\bibnamefont{{Donovan}}},
  \bibinfo{author}{\bibfnamefont{C.~J.} \bibnamefont{{Ma}}},
  \bibinfo{author}{\bibfnamefont{A.~C.} \bibnamefont{{Edge}}},
  \bibnamefont{and} \bibinfo{author}{\bibfnamefont{L.}~\bibnamefont{{van
  Speybroeck}}}, \bibinfo{journal}{\apjl} \textbf{\bibinfo{volume}{661}},
  \bibinfo{pages}{L33} (\bibinfo{year}{2007}), \eprint{astro-ph/0703394}.

\bibitem[{\citenamefont{{Zitrin} and {Broadhurst}}(2009)}]{2009ApJ...703L.132Z}
\bibinfo{author}{\bibfnamefont{A.}~\bibnamefont{{Zitrin}}} \bibnamefont{and}
  \bibinfo{author}{\bibfnamefont{T.}~\bibnamefont{{Broadhurst}}},
  \bibinfo{journal}{\apjl} \textbf{\bibinfo{volume}{703}},
  \bibinfo{pages}{L132} (\bibinfo{year}{2009}), \eprint{0906.5079}.

\bibitem[{\citenamefont{{Smith} et~al.}(2009)\citenamefont{{Smith}, {Ebeling},
  {Limousin}, {Kneib}, {Swinbank}, {Ma}, {Jauzac}, {Richard}, {Jullo}, {Sand}
  et~al.}}]{2009ApJ...707L.163S}
\bibinfo{author}{\bibfnamefont{G.~P.} \bibnamefont{{Smith}}},
  \bibinfo{author}{\bibfnamefont{H.}~\bibnamefont{{Ebeling}}},
  \bibinfo{author}{\bibfnamefont{M.}~\bibnamefont{{Limousin}}},
  \bibinfo{author}{\bibfnamefont{J.-P.} \bibnamefont{{Kneib}}},
  \bibinfo{author}{\bibfnamefont{A.~M.} \bibnamefont{{Swinbank}}},
  \bibinfo{author}{\bibfnamefont{C.-J.} \bibnamefont{{Ma}}},
  \bibinfo{author}{\bibfnamefont{M.}~\bibnamefont{{Jauzac}}},
  \bibinfo{author}{\bibfnamefont{J.}~\bibnamefont{{Richard}}},
  \bibinfo{author}{\bibfnamefont{E.}~\bibnamefont{{Jullo}}},
  \bibinfo{author}{\bibfnamefont{D.~J.} \bibnamefont{{Sand}}},
  \bibnamefont{et~al.}, \bibinfo{journal}{\apjl}
  \textbf{\bibinfo{volume}{707}}, \bibinfo{pages}{L163} (\bibinfo{year}{2009}),
  \eprint{0911.2003}.

\bibitem[{\citenamefont{{Zheng} et~al.}(2012)\citenamefont{{Zheng}, {Postman},
  {Zitrin}, {Moustakas}, {Shu}, {Jouvel}, {H{\o}st}, {Molino}, {Bradley}, {Coe}
  et~al.}}]{2012Natur.489..406Z}
\bibinfo{author}{\bibfnamefont{W.}~\bibnamefont{{Zheng}}},
  \bibinfo{author}{\bibfnamefont{M.}~\bibnamefont{{Postman}}},
  \bibinfo{author}{\bibfnamefont{A.}~\bibnamefont{{Zitrin}}},
  \bibinfo{author}{\bibfnamefont{J.}~\bibnamefont{{Moustakas}}},
  \bibinfo{author}{\bibfnamefont{X.}~\bibnamefont{{Shu}}},
  \bibinfo{author}{\bibfnamefont{S.}~\bibnamefont{{Jouvel}}},
  \bibinfo{author}{\bibfnamefont{O.}~\bibnamefont{{H{\o}st}}},
  \bibinfo{author}{\bibfnamefont{A.}~\bibnamefont{{Molino}}},
  \bibinfo{author}{\bibfnamefont{L.}~\bibnamefont{{Bradley}}},
  \bibinfo{author}{\bibfnamefont{D.}~\bibnamefont{{Coe}}},
  \bibnamefont{et~al.}, \bibinfo{journal}{\nat} \textbf{\bibinfo{volume}{489}},
  \bibinfo{pages}{406} (\bibinfo{year}{2012}), \eprint{1204.2305}.

\bibitem[{\citenamefont{{Rau} et~al.}(2014)\citenamefont{{Rau}, {Vegetti}, and
  {White}}}]{2014MNRAS.443..957R}
\bibinfo{author}{\bibfnamefont{S.}~\bibnamefont{{Rau}}},
  \bibinfo{author}{\bibfnamefont{S.}~\bibnamefont{{Vegetti}}},
  \bibnamefont{and} \bibinfo{author}{\bibfnamefont{S.~D.~M.}
  \bibnamefont{{White}}}, \bibinfo{journal}{\mnras}
  \textbf{\bibinfo{volume}{443}}, \bibinfo{pages}{957} (\bibinfo{year}{2014}),
  \eprint{1402.7321}.

\bibitem[{\citenamefont{{Richard} et~al.}(2014)\citenamefont{{Richard},
  {Jauzac}, {Limousin}, {Jullo}, {Cl{\'e}ment}, {Ebeling}, {Kneib}, {Atek},
  {Natarajan}, {Egami} et~al.}}]{2014MNRAS.444..268R}
\bibinfo{author}{\bibfnamefont{J.}~\bibnamefont{{Richard}}},
  \bibinfo{author}{\bibfnamefont{M.}~\bibnamefont{{Jauzac}}},
  \bibinfo{author}{\bibfnamefont{M.}~\bibnamefont{{Limousin}}},
  \bibinfo{author}{\bibfnamefont{E.}~\bibnamefont{{Jullo}}},
  \bibinfo{author}{\bibfnamefont{B.}~\bibnamefont{{Cl{\'e}ment}}},
  \bibinfo{author}{\bibfnamefont{H.}~\bibnamefont{{Ebeling}}},
  \bibinfo{author}{\bibfnamefont{J.-P.} \bibnamefont{{Kneib}}},
  \bibinfo{author}{\bibfnamefont{H.}~\bibnamefont{{Atek}}},
  \bibinfo{author}{\bibfnamefont{P.}~\bibnamefont{{Natarajan}}},
  \bibinfo{author}{\bibfnamefont{E.}~\bibnamefont{{Egami}}},
  \bibnamefont{et~al.}, \bibinfo{journal}{\mnras}
  \textbf{\bibinfo{volume}{444}}, \bibinfo{pages}{268} (\bibinfo{year}{2014}),
  \eprint{1405.3303}.

\bibitem[{\citenamefont{{Jauzac} et~al.}(2016)\citenamefont{{Jauzac},
  {Richard}, {Limousin}, {Knowles}, {Mahler}, {Smith}, {Kneib}, {Jullo},
  {Natarajan}, {Ebeling} et~al.}}]{2016MNRAS.457.2029J}
\bibinfo{author}{\bibfnamefont{M.}~\bibnamefont{{Jauzac}}},
  \bibinfo{author}{\bibfnamefont{J.}~\bibnamefont{{Richard}}},
  \bibinfo{author}{\bibfnamefont{M.}~\bibnamefont{{Limousin}}},
  \bibinfo{author}{\bibfnamefont{K.}~\bibnamefont{{Knowles}}},
  \bibinfo{author}{\bibfnamefont{G.}~\bibnamefont{{Mahler}}},
  \bibinfo{author}{\bibfnamefont{G.~P.} \bibnamefont{{Smith}}},
  \bibinfo{author}{\bibfnamefont{J.~P.} \bibnamefont{{Kneib}}},
  \bibinfo{author}{\bibfnamefont{E.}~\bibnamefont{{Jullo}}},
  \bibinfo{author}{\bibfnamefont{P.}~\bibnamefont{{Natarajan}}},
  \bibinfo{author}{\bibfnamefont{H.}~\bibnamefont{{Ebeling}}},
  \bibnamefont{et~al.}, \bibinfo{journal}{\mnras}
  \textbf{\bibinfo{volume}{457}}, \bibinfo{pages}{2029} (\bibinfo{year}{2016}),
  \eprint{1509.08914}.

\bibitem[{\citenamefont{{Treu} et~al.}(2016)\citenamefont{{Treu}, {Brammer},
  {Diego}, {Grillo}, {Kelly}, {Oguri}, {Rodney}, {Rosati}, {Sharon}, {Zitrin}
  et~al.}}]{2016ApJ...817...60T}
\bibinfo{author}{\bibfnamefont{T.}~\bibnamefont{{Treu}}},
  \bibinfo{author}{\bibfnamefont{G.}~\bibnamefont{{Brammer}}},
  \bibinfo{author}{\bibfnamefont{J.~M.} \bibnamefont{{Diego}}},
  \bibinfo{author}{\bibfnamefont{C.}~\bibnamefont{{Grillo}}},
  \bibinfo{author}{\bibfnamefont{P.~L.} \bibnamefont{{Kelly}}},
  \bibinfo{author}{\bibfnamefont{M.}~\bibnamefont{{Oguri}}},
  \bibinfo{author}{\bibfnamefont{S.~A.} \bibnamefont{{Rodney}}},
  \bibinfo{author}{\bibfnamefont{P.}~\bibnamefont{{Rosati}}},
  \bibinfo{author}{\bibfnamefont{K.}~\bibnamefont{{Sharon}}},
  \bibinfo{author}{\bibfnamefont{A.}~\bibnamefont{{Zitrin}}},
  \bibnamefont{et~al.}, \bibinfo{journal}{\apj} \textbf{\bibinfo{volume}{817}},
  \bibinfo{eid}{60} (\bibinfo{year}{2016}), \eprint{1510.05750}.

\bibitem[{\citenamefont{{Grillo} et~al.}(2016)\citenamefont{{Grillo}, {Karman},
  {Suyu}, {Rosati}, {Balestra}, {Mercurio}, {Lombardi}, {Treu}, {Caminha},
  {Halkola} et~al.}}]{2016ApJ...822...78G}
\bibinfo{author}{\bibfnamefont{C.}~\bibnamefont{{Grillo}}},
  \bibinfo{author}{\bibfnamefont{W.}~\bibnamefont{{Karman}}},
  \bibinfo{author}{\bibfnamefont{S.~H.} \bibnamefont{{Suyu}}},
  \bibinfo{author}{\bibfnamefont{P.}~\bibnamefont{{Rosati}}},
  \bibinfo{author}{\bibfnamefont{I.}~\bibnamefont{{Balestra}}},
  \bibinfo{author}{\bibfnamefont{A.}~\bibnamefont{{Mercurio}}},
  \bibinfo{author}{\bibfnamefont{M.}~\bibnamefont{{Lombardi}}},
  \bibinfo{author}{\bibfnamefont{T.}~\bibnamefont{{Treu}}},
  \bibinfo{author}{\bibfnamefont{G.~B.} \bibnamefont{{Caminha}}},
  \bibinfo{author}{\bibfnamefont{A.}~\bibnamefont{{Halkola}}},
  \bibnamefont{et~al.}, \bibinfo{journal}{\apj} \textbf{\bibinfo{volume}{822}},
  \bibinfo{eid}{78} (\bibinfo{year}{2016}), \eprint{1511.04093}.

\bibitem[{\citenamefont{{Brammer} et~al.}(2016)\citenamefont{{Brammer},
  {Marchesini}, {Labb{\'e}}, {Spitler}, {Lange-Vagle}, {Barker}, {Tanaka},
  {Fontana}, {Galametz}, {Ferr{\'e}-Mateu} et~al.}}]{2016ApJS..226....6B}
\bibinfo{author}{\bibfnamefont{G.~B.} \bibnamefont{{Brammer}}},
  \bibinfo{author}{\bibfnamefont{D.}~\bibnamefont{{Marchesini}}},
  \bibinfo{author}{\bibfnamefont{I.}~\bibnamefont{{Labb{\'e}}}},
  \bibinfo{author}{\bibfnamefont{L.}~\bibnamefont{{Spitler}}},
  \bibinfo{author}{\bibfnamefont{D.}~\bibnamefont{{Lange-Vagle}}},
  \bibinfo{author}{\bibfnamefont{E.~A.} \bibnamefont{{Barker}}},
  \bibinfo{author}{\bibfnamefont{M.}~\bibnamefont{{Tanaka}}},
  \bibinfo{author}{\bibfnamefont{A.}~\bibnamefont{{Fontana}}},
  \bibinfo{author}{\bibfnamefont{A.}~\bibnamefont{{Galametz}}},
  \bibinfo{author}{\bibfnamefont{A.}~\bibnamefont{{Ferr{\'e}-Mateu}}},
  \bibnamefont{et~al.}, \bibinfo{journal}{\apjs}
  \textbf{\bibinfo{volume}{226}}, \bibinfo{eid}{6} (\bibinfo{year}{2016}),
  \eprint{1606.07450}.

\bibitem[{\citenamefont{{Oguri}}(2010)}]{2010PASJ...62.1017O}
\bibinfo{author}{\bibfnamefont{M.}~\bibnamefont{{Oguri}}},
  \bibinfo{journal}{\pasj} \textbf{\bibinfo{volume}{62}}, \bibinfo{pages}{1017}
  (\bibinfo{year}{2010}), \eprint{1005.3103}.

\bibitem[{\citenamefont{{Oguri}}(2021)}]{2021PASP..133g4504O}
\bibinfo{author}{\bibfnamefont{M.}~\bibnamefont{{Oguri}}},
  \bibinfo{journal}{\pasp} \textbf{\bibinfo{volume}{133}},
  \bibinfo{eid}{074504} (\bibinfo{year}{2021}), \eprint{2106.11464}.

\bibitem[{\citenamefont{{Navarro} et~al.}(1997)\citenamefont{{Navarro},
  {Frenk}, and {White}}}]{1997ApJ...490..493N}
\bibinfo{author}{\bibfnamefont{J.~F.} \bibnamefont{{Navarro}}},
  \bibinfo{author}{\bibfnamefont{C.~S.} \bibnamefont{{Frenk}}},
  \bibnamefont{and} \bibinfo{author}{\bibfnamefont{S.~D.~M.}
  \bibnamefont{{White}}}, \bibinfo{journal}{\apj}
  \textbf{\bibinfo{volume}{490}}, \bibinfo{pages}{493} (\bibinfo{year}{1997}),
  \eprint{astro-ph/9611107}.

\bibitem[{\citenamefont{{Jaffe}}(1983)}]{1983MNRAS.202..995J}
\bibinfo{author}{\bibfnamefont{W.}~\bibnamefont{{Jaffe}}},
  \bibinfo{journal}{\mnras} \textbf{\bibinfo{volume}{202}},
  \bibinfo{pages}{995} (\bibinfo{year}{1983}).

\bibitem[{\citenamefont{{Keeton}}(2001)}]{2001astro.ph..2341K}
\bibinfo{author}{\bibfnamefont{C.~R.} \bibnamefont{{Keeton}}},
  \bibinfo{journal}{arXiv e-prints} \bibinfo{eid}{astro-ph/0102341}
  (\bibinfo{year}{2001}), \eprint{astro-ph/0102341}.

\bibitem[{\citenamefont{{Evans} and {Witt}}(2003)}]{2003MNRAS.345.1351E}
\bibinfo{author}{\bibfnamefont{N.~W.} \bibnamefont{{Evans}}} \bibnamefont{and}
  \bibinfo{author}{\bibfnamefont{H.~J.} \bibnamefont{{Witt}}},
  \bibinfo{journal}{\mnras} \textbf{\bibinfo{volume}{345}},
  \bibinfo{pages}{1351} (\bibinfo{year}{2003}), \eprint{astro-ph/0212013}.

\bibitem[{\citenamefont{{Kawano} et~al.}(2004)\citenamefont{{Kawano}, {Oguri},
  {Matsubara}, and {Ikeuchi}}}]{2004PASJ...56..253K}
\bibinfo{author}{\bibfnamefont{Y.}~\bibnamefont{{Kawano}}},
  \bibinfo{author}{\bibfnamefont{M.}~\bibnamefont{{Oguri}}},
  \bibinfo{author}{\bibfnamefont{T.}~\bibnamefont{{Matsubara}}},
  \bibnamefont{and}
  \bibinfo{author}{\bibfnamefont{S.}~\bibnamefont{{Ikeuchi}}},
  \bibinfo{journal}{\pasj} \textbf{\bibinfo{volume}{56}}, \bibinfo{pages}{253}
  (\bibinfo{year}{2004}), \eprint{astro-ph/0404013}.

\bibitem[{\citenamefont{{Congdon} and {Keeton}}(2005)}]{2005MNRAS.364.1459C}
\bibinfo{author}{\bibfnamefont{A.~B.} \bibnamefont{{Congdon}}}
  \bibnamefont{and} \bibinfo{author}{\bibfnamefont{C.~R.}
  \bibnamefont{{Keeton}}}, \bibinfo{journal}{\mnras}
  \textbf{\bibinfo{volume}{364}}, \bibinfo{pages}{1459} (\bibinfo{year}{2005}),
  \eprint{astro-ph/0510232}.

\bibitem[{\citenamefont{{Yoo} et~al.}(2006)\citenamefont{{Yoo}, {Kochanek},
  {Falco}, and {McLeod}}}]{2006ApJ...642...22Y}
\bibinfo{author}{\bibfnamefont{J.}~\bibnamefont{{Yoo}}},
  \bibinfo{author}{\bibfnamefont{C.~S.} \bibnamefont{{Kochanek}}},
  \bibinfo{author}{\bibfnamefont{E.~E.} \bibnamefont{{Falco}}},
  \bibnamefont{and} \bibinfo{author}{\bibfnamefont{B.~A.}
  \bibnamefont{{McLeod}}}, \bibinfo{journal}{\apj}
  \textbf{\bibinfo{volume}{642}}, \bibinfo{pages}{22} (\bibinfo{year}{2006}),
  \eprint{astro-ph/0511001}.

\bibitem[{\citenamefont{{Oguri} et~al.}(2013)\citenamefont{{Oguri},
  {Schrabback}, {Jullo}, {Ota}, {Kochanek}, {Dai}, {Ofek}, {Richards},
  {Blandford}, {Falco} et~al.}}]{2013MNRAS.429..482O}
\bibinfo{author}{\bibfnamefont{M.}~\bibnamefont{{Oguri}}},
  \bibinfo{author}{\bibfnamefont{T.}~\bibnamefont{{Schrabback}}},
  \bibinfo{author}{\bibfnamefont{E.}~\bibnamefont{{Jullo}}},
  \bibinfo{author}{\bibfnamefont{N.}~\bibnamefont{{Ota}}},
  \bibinfo{author}{\bibfnamefont{C.~S.} \bibnamefont{{Kochanek}}},
  \bibinfo{author}{\bibfnamefont{X.}~\bibnamefont{{Dai}}},
  \bibinfo{author}{\bibfnamefont{E.~O.} \bibnamefont{{Ofek}}},
  \bibinfo{author}{\bibfnamefont{G.~T.} \bibnamefont{{Richards}}},
  \bibinfo{author}{\bibfnamefont{R.~D.} \bibnamefont{{Blandford}}},
  \bibinfo{author}{\bibfnamefont{E.~E.} \bibnamefont{{Falco}}},
  \bibnamefont{et~al.}, \bibinfo{journal}{\mnras}
  \textbf{\bibinfo{volume}{429}}, \bibinfo{pages}{482} (\bibinfo{year}{2013}),
  \eprint{1209.0458}.

\bibitem[{\citenamefont{{Keeton} et~al.}(1997)\citenamefont{{Keeton},
  {Kochanek}, and {Seljak}}}]{1997ApJ...482..604K}
\bibinfo{author}{\bibfnamefont{C.~R.} \bibnamefont{{Keeton}}},
  \bibinfo{author}{\bibfnamefont{C.~S.} \bibnamefont{{Kochanek}}},
  \bibnamefont{and} \bibinfo{author}{\bibfnamefont{U.}~\bibnamefont{{Seljak}}},
  \bibinfo{journal}{\apj} \textbf{\bibinfo{volume}{482}}, \bibinfo{pages}{604}
  (\bibinfo{year}{1997}), \eprint{astro-ph/9610163}.

\bibitem[{\citenamefont{{Birrer} and {Treu}}(2019)}]{2019MNRAS.489.2097B}
\bibinfo{author}{\bibfnamefont{S.}~\bibnamefont{{Birrer}}} \bibnamefont{and}
  \bibinfo{author}{\bibfnamefont{T.}~\bibnamefont{{Treu}}},
  \bibinfo{journal}{\mnras} \textbf{\bibinfo{volume}{489}},
  \bibinfo{pages}{2097} (\bibinfo{year}{2019}), \eprint{1904.10965}.

\bibitem[{\citenamefont{{Witt} et~al.}(1995)\citenamefont{{Witt}, {Mao}, and
  {Schechter}}}]{1995ApJ...443...18W}
\bibinfo{author}{\bibfnamefont{H.~J.} \bibnamefont{{Witt}}},
  \bibinfo{author}{\bibfnamefont{S.}~\bibnamefont{{Mao}}}, \bibnamefont{and}
  \bibinfo{author}{\bibfnamefont{P.~L.} \bibnamefont{{Schechter}}},
  \bibinfo{journal}{\apj} \textbf{\bibinfo{volume}{443}}, \bibinfo{pages}{18}
  (\bibinfo{year}{1995}).

\bibitem[{\citenamefont{{Kochanek}}(2002)}]{2002ApJ...578...25K}
\bibinfo{author}{\bibfnamefont{C.~S.} \bibnamefont{{Kochanek}}},
  \bibinfo{journal}{\apj} \textbf{\bibinfo{volume}{578}}, \bibinfo{pages}{25}
  (\bibinfo{year}{2002}), \eprint{astro-ph/0205319}.

\bibitem[{\citenamefont{{Oguri} and {Kawano}}(2003)}]{2003MNRAS.338L..25O}
\bibinfo{author}{\bibfnamefont{M.}~\bibnamefont{{Oguri}}} \bibnamefont{and}
  \bibinfo{author}{\bibfnamefont{Y.}~\bibnamefont{{Kawano}}},
  \bibinfo{journal}{\mnras} \textbf{\bibinfo{volume}{338}},
  \bibinfo{pages}{L25} (\bibinfo{year}{2003}), \eprint{astro-ph/0211499}.

\bibitem[{\citenamefont{{Frye} et~al.}(2024)\citenamefont{{Frye}, {Pascale},
  {Pierel}, {Chen}, {Foo}, {Leimbach}, {Garuda}, {Cohen}, {Kamieneski},
  {Windhorst} et~al.}}]{2024ApJ...961..171F}
\bibinfo{author}{\bibfnamefont{B.~L.} \bibnamefont{{Frye}}},
  \bibinfo{author}{\bibfnamefont{M.}~\bibnamefont{{Pascale}}},
  \bibinfo{author}{\bibfnamefont{J.}~\bibnamefont{{Pierel}}},
  \bibinfo{author}{\bibfnamefont{W.}~\bibnamefont{{Chen}}},
  \bibinfo{author}{\bibfnamefont{N.}~\bibnamefont{{Foo}}},
  \bibinfo{author}{\bibfnamefont{R.}~\bibnamefont{{Leimbach}}},
  \bibinfo{author}{\bibfnamefont{N.}~\bibnamefont{{Garuda}}},
  \bibinfo{author}{\bibfnamefont{S.~H.} \bibnamefont{{Cohen}}},
  \bibinfo{author}{\bibfnamefont{P.~S.} \bibnamefont{{Kamieneski}}},
  \bibinfo{author}{\bibfnamefont{R.~A.} \bibnamefont{{Windhorst}}},
  \bibnamefont{et~al.}, \bibinfo{journal}{\apj} \textbf{\bibinfo{volume}{961}},
  \bibinfo{eid}{171} (\bibinfo{year}{2024}), \eprint{2309.07326}.

\end{thebibliography}

\end{document}